\documentclass[12pt]{article}

\usepackage{epsfig}

\def\be{\begin{equation}}
\def\ee{\end{equation}}
\def\bea{\begin{eqnarray}}
\def\eea{\end{eqnarray}}

\def\[{\lfloor{\hskip 0.35pt}\!\!\!\lceil}
\def\]{\rfloor{\hskip 0.35pt}\!\!\!\rceil}

\def\IR{\relax{\rm I\kern-.18em R}}
\def\IC{\relax{\rm I\kern-.18em C}}

\newcommand{\AmS}{{\protect\the\textfont2
   A\kern-.1667em\lower.5ex\hbox{M}\kern-.125emS}}

\catcode`@=11
\def\un#1{\relax\ifmmode\@@underline#1\else
         $\@@underline{\hbox{#1}}$\relax\fi}
\catcode`@=12

\def\fracm#1#2{\hbox{\large{${\frac{{#1}}{{#2}}}$}}}

\def\ad{{\kern0.5pt
                    \alpha \kern-5.05pt
\raise5.8pt\hbox{$\textstyle.$}\kern
0.5pt}}

\def\Dot#1{{\kern0.5pt
      {#1} \kern-5.05pt \raise5.8pt\hbox{$\textstyle.$}\kern
0.5pt}}

\def\a{\alpha}
\def\b{\beta}

\def\d{\delta}
\def\e{\epsilon}

\def\g{\gamma}

\def\k{\kappa}
\def\l{\lambda}
\def\m{\mu}
\def\n{\nu}
\def\o{\omega}

\def\r{\rho}
\def\s{\sigma}
\def\t{\tau}

\def\z{\zeta}
\def\D{\Delta}

\def\G{\Gamma}

\def\L{\Lambda}

\def\P{\Pi}

\def\S{\Sigma}
\def\U{\Upsilon}

\def\bo{{\raise.15ex\hbox{\large$\Box$}}}               
\def\pa{\partial}                                       
\def\de{\nabla}                                         
\def\TH{{\raise.2ex\hbox{$\displaystyle \bigodot$}\mskip-4.7mu \llap H
\;}}
\def\face{{\raise.2ex\hbox{$\displaystyle \bigodot$}\mskip-2.2mu \llap
{$\ddot
         \smile$}}}                                      

\def\Hat#1{\widehat{#1}}                        
\def\Bar#1{\overline{#1}}                       
\def\VEV#1{\left\langle #1\right\rangle}        
\def\leftrightarrowfill{$\mathsurround=0pt \mathord\leftarrow \mkern-6mu
         \cleaders\hbox{$\mkern-2mu \mathord- \mkern-2mu$}\hfill
         \mkern-6mu \mathord\rightarrow$}
\def\dvec#1{\vbox{\ialign{##\crcr
         \leftrightarrowfill\crcr\noalign{\kern-1pt\nointerlineskip}
         $\hfil\displaystyle{#1}\hfil$\crcr}}}           

\def\fracm#1#2{\hbox{\large{${\frac{{#1}}{{#2}}}$}}}
\def\frac#1#2{{\textstyle{#1\over\vphantom2\smash{\raise.20ex
         \hbox{$\scriptstyle{#2}$}}}}}                   
\def\ha{\frac12}                                        
\def\sfrac#1#2{{\vphantom1\smash{\lower.5ex\hbox{\small$#1$}}\over
         \vphantom1\smash{\raise.4ex\hbox{\small$#2$}}}} 
\def\bfrac#1#2{{\vphantom1\smash{\lower.5ex\hbox{$#1$}}\over
         \vphantom1\smash{\raise.3ex\hbox{$#2$}}}}       
\def\afrac#1#2{{\vphantom1\smash{\lower.5ex\hbox{$#1$}}\over#2}}    
\def\parvar#1#2{{\d #1\over \d #2}}               

\newskip\humongous \humongous=0pt plus 1000pt minus 1000pt
\def\caja{\mathsurround=0pt}
\def\eqalign#1{\,\vcenter{\openup2\jot \caja
         \ialign{\strut \hfil$\displaystyle{##}$&$
         \displaystyle{{}##}$\hfil\crcr#1\crcr}}\,}
\newif\ifdtup

   \def\pp{{\mathchoice
           {
               \kern 1pt%
               \raise 1pt
               \vbox{\hrule width5pt height0.4pt depth0pt
                     \kern -2pt
                     \hbox{\kern 2.3pt
                           \vrule width0.4pt height6pt depth0pt
                           }
                     \kern -2pt
                     \hrule width5pt height0.4pt depth0pt}%
                     \kern 1pt
            }
             {
               \kern 1pt%
               \raise 1pt
               \vbox{\hrule width4.3pt height0.4pt depth0pt
                     \kern -1.8pt
                     \hbox{\kern 1.95pt
                           \vrule width0.4pt height5.4pt depth0pt
                           }
                     \kern -1.8pt
                     \hrule width4.3pt height0.4pt depth0pt}%
                     \kern 1pt
             }
             {
               \kern 0.5pt%
               \raise 1pt
               \vbox{\hrule width4.0pt height0.3pt depth0pt
                     \kern -1.9pt  
                     \hbox{\kern 1.85pt
                           \vrule width0.3pt height5.7pt depth0pt
                           }
                     \kern -1.9pt
                     \hrule width4.0pt height0.3pt depth0pt}%
                     \kern 0.5pt
             }
             {
               \kern 0.5pt%
               \raise 1pt
               \vbox{\hrule width3.6pt height0.3pt depth0pt
                     \kern -1.5pt
                     \hbox{\kern 1.65pt
                           \vrule width0.3pt height4.5pt depth0pt
                           }
                     \kern -1.5pt
                     \hrule width3.6pt height0.3pt depth0pt}%
                     \kern 0.5pt
             }
         }}

   \def\mm{{\mathchoice
                        {
                              \kern 1pt
                \raise 1pt    \vbox{\hrule width5pt height0.4pt depth0pt
                                   \kern 2pt
                                   \hrule width5pt height0.4pt depth0pt}
                              \kern 1pt}
                        {
                             \kern 1pt
                \raise 1pt \vbox{\hrule width4.3pt height0.4pt depth0pt
                                   \kern 1.8pt
                                   \hrule width4.3pt height0.4pt depth0pt}
                              \kern 1pt}
                        {
                             \kern 0.5pt
                \raise 1pt
                             \vbox{\hrule width4.0pt height0.3pt depth0pt
                                   \kern 1.9pt
                                   \hrule width4.0pt height0.3pt depth0pt}
                             \kern 1pt}
                        {
                            \kern 0.5pt
              \raise 1pt  \vbox{\hrule width3.6pt height0.3pt depth0pt
                                   \kern 1.5pt
                                   \hrule width3.6pt height0.3pt depth0pt}
                            \kern 0.5pt}
                        }}

\def\pd{{\kern0.5pt
                    + \kern-5.05pt \raise5.8pt\hbox{$\textstyle.$}\kern
0.5pt}}

\def\pmd{{\kern0.5pt
                   \pm \kern-5.05pt \raise6.3pt\hbox{$\textstyle.$}\kern1.5pt}}

\def\md{{\mathchoice
    {
       {{\kern 1pt - \kern-6.2pt \raise5pt\hbox{$\textstyle.$}\kern 1pt}}}
     {
       {{\kern 1pt - \kern-6.2pt \raise5pt\hbox{$\textstyle.$}\kern 1pt}}}
     {
       {\kern0.5pt - \kern-5.05pt \raise3.4pt\hbox{$\textstyle.$}\kern0.5pt}}
     {
       {\kern0.5pt - \kern-5.05pt \raise3.4pt\hbox{$\textstyle.$}\kern0.5pt}}}}

\def\ad{{\dot{\alpha}}}

\def\pp{{\mathchoice
           {
               \kern 1pt%
               \raise 1pt
               \vbox{\hrule width5pt height0.4pt depth0pt
                     \kern -2pt
                     \hbox{\kern 2.3pt
                           \vrule width0.4pt height6pt depth0pt
                           }
                     \kern -2pt
                     \hrule width5pt height0.4pt depth0pt}%
                     \kern 1pt
            }
             {
               \kern 1pt%
               \raise 1pt
               \vbox{\hrule width4.3pt height0.4pt depth0pt
                     \kern -1.8pt
                     \hbox{\kern 1.95pt
                           \vrule width0.4pt height5.4pt depth0pt
                           }
                     \kern -1.8pt
                     \hrule width4.3pt height0.4pt depth0pt}%
                     \kern 1pt
             }
             {
               \kern 0.5pt%
               \raise 1pt
               \vbox{\hrule width4.0pt height0.3pt depth0pt
                     \kern -1.9pt  
                     \hbox{\kern 1.85pt
                           \vrule width0.3pt height5.7pt depth0pt
                           }
                     \kern -1.9pt
                     \hrule width4.0pt height0.3pt depth0pt}%
                     \kern 0.5pt
             }
             {
               \kern 0.5pt%
               \raise 1pt
               \vbox{\hrule width3.6pt height0.3pt depth0pt
                     \kern -1.5pt
                     \hbox{\kern 1.65pt
                           \vrule width0.3pt height4.5pt depth0pt
                           }
                     \kern -1.5pt
                     \hrule width3.6pt height0.3pt depth0pt}%
                     \kern 0.5pt
             }
         }}

   \def\mm{{\mathchoice
                        {
                              \kern 1pt
                \raise 1pt    \vbox{\hrule width5pt height0.4pt depth0pt
                                   \kern 2pt
                                   \hrule width5pt height0.4pt depth0pt}
                              \kern 1pt}
                        {
                             \kern 1pt
                \raise 1pt \vbox{\hrule width4.3pt height0.4pt depth0pt
                                   \kern 1.8pt
                                   \hrule width4.3pt height0.4pt depth0pt}
                              \kern 1pt}
                        {
                             \kern 0.5pt
                \raise 1pt
                             \vbox{\hrule width4.0pt height0.3pt depth0pt
                                   \kern 1.9pt
                                   \hrule width4.0pt height0.3pt depth0pt}
                             \kern 1pt}
                        {
                            \kern 0.5pt
              \raise 1pt  \vbox{\hrule width3.6pt height0.3pt depth0pt
                                   \kern 1.5pt
                                   \hrule width3.6pt height0.3pt depth0pt}
                            \kern 0.5pt}
                        }}

\def\pd{{\kern0.5pt
                    + \kern-5.05pt \raise5.8pt\hbox{$\textstyle.$}\kern
0.5pt}}

\def\pmd{{\kern0.5pt
                   \pm \kern-5.05pt \raise6.3pt\hbox{$\textstyle.$}\kern1.5pt}}

\def\md{{\mathchoice
    {
       {{\kern 1pt - \kern-6.2pt \raise5pt\hbox{$\textstyle.$}\kern 1pt}}}
     {
       {{\kern 1pt - \kern-6.2pt \raise5pt\hbox{$\textstyle.$}\kern 1pt}}}
     {
       {\kern0.5pt - \kern-5.05pt \raise3.4pt\hbox{$\textstyle.$}\kern0.5pt}}
     {
       {\kern0.5pt - \kern-5.05pt \raise3.4pt\hbox{$\textstyle.$}\kern0.5pt}}}}

\def\dslash{\not{\hbox{\kern-2pt $\partial$}}}
\def\Dslash{\not{\hbox{\kern-4pt $D$}}}
\def\pslash{\not{\hbox{\kern-2.3pt $p$}}}
  \newtoks\slashfraction
  \slashfraction={.13}
  \def\slash#1{\setbox0\hbox{$ #1 $}
  \setbox0\hbox to \the\slashfraction\wd0{\hss \box0}/\box0 }

\def\Sc#1{{\hbox{\sc #1}}}      
\font\ro=cmsy10                          
\def\kcr{{\hbox{\ro \char'170}}}                
\def\ktl{{\hbox{\ro \char'170}}}        
\def\ktr{{\hbox{\ro \char'170}}}        
\def\kbl{{\hbox{\ro \char'170}}}        
\def\kbr{{\hbox{\ro \char'170}}}        

\def\plpl{\raise-2pt\hbox{$\raise3pt\hbox{$_+$}\hskip-6.67pt\raise0.0pt
\hbox{$^+$}\hskip 0.01pt$}}
\def\mimi{\raise-2pt\hbox{$\raise3pt\hbox{$_-$}\hskip-6.67pt\raise0.0pt
\hbox{$^-$}\hskip 0.01pt$}}

\def\bo{{\raise.15ex\hbox{\large$\Box$}}}               
\def\pa{\partial}                                       
\def\de{\nabla}                                         
\def\TH{{\raise.2ex\hbox{$\displaystyle \bigodot$}\mskip-4.7mu \llap H \;}}
\def\face{{\raise.2ex\hbox{$\displaystyle \bigodot$}\mskip-2.2mu \llap {$\ddot
         \smile$}}}                                      

\def\Hat#1{\widehat{#1}}                        
\def\Bar#1{\overline{#1}}                       
\def\VEV#1{\left\langle #1\right\rangle}        
\def\leftrightarrowfill{$\mathsurround=0pt \mathord\leftarrow \mkern-6mu
         \cleaders\hbox{$\mkern-2mu \mathord- \mkern-2mu$}\hfill
         \mkern-6mu \mathord\rightarrow$}
\def\dvec#1{\vbox{\ialign{##\crcr
         \leftrightarrowfill\crcr\noalign{\kern-1pt\nointerlineskip}
         $\hfil\displaystyle{#1}\hfil$\crcr}}}           

\def\fracm#1#2{\hbox{\large{${\frac{{#1}}{{#2}}}$}}}
\def\frac#1#2{{\textstyle{#1\over\vphantom2\smash{\raise.20ex
         \hbox{$\scriptstyle{#2}$}}}}}                   
\def\ha{\frac12}                                        
\def\sfrac#1#2{{\vphantom1\smash{\lower.5ex\hbox{\small$#1$}}\over
         \vphantom1\smash{\raise.4ex\hbox{\small$#2$}}}} 
\def\bfrac#1#2{{\vphantom1\smash{\lower.5ex\hbox{$#1$}}\over
         \vphantom1\smash{\raise.3ex\hbox{$#2$}}}}       
\def\afrac#1#2{{\vphantom1\smash{\lower.5ex\hbox{$#1$}}\over#2}}    
\def\parvar#1#2{{\d #1\over \d #2}}               

\parskip=\medskipamount                 
\thicklines                         

\def\oldheadpic{                                
         \setlength{\unitlength}{.4mm}
         \thinlines
         \par
         \begin{picture}(349,16)
         \put(325,16){\line(1,0){4}}
         \put(330,16){\line(1,0){4}}
         \put(340,16){\line(1,0){4}}
         \put(335,0){\line(1,0){4}}
         \put(340,0){\line(1,0){4}}
         \put(345,0){\line(1,0){4}}
         \put(329,0){\line(0,1){16}}
         \put(330,0){\line(0,1){16}}
         \put(339,0){\line(0,1){16}}
         \put(340,0){\line(0,1){16}}
         \put(344,0){\line(0,1){16}}
         \put(345,0){\line(0,1){16}}
         \put(329,16){\oval(8,32)[bl]}
         \put(330,16){\oval(8,32)[br]}
         \put(339,0){\oval(8,32)[tl]}
         \put(345,0){\oval(8,32)[tr]}
         \end{picture}
         \par
         \thicklines
         \vskip.2in}
\def\oldtitle#1#2#3#4{\oldheadpic\begin{center}\vglue.5in{\large\bf #1}\\[.6in]
         {#2}\\[.1in] {\it Department of Physics and Astronomy}\\
         {\it University of Maryland, College Park, MD 20742}\\[.6in]
         Physics Publication \#{#3}\\ {#4}\\[1.5in] {\bf ABSTRACT}\\[.1in]
         \end{center} \begin{quotation}}                 
\def\oldTitle#1#2#3#4#5#6#7{\oldheadpic\begin{center} \vglue .4in
         {\large\bf #1}\\[.4in]
         {#2}\\[.1in] {\it Department of Physics and Astronomy}\\
         {\it University of Maryland, College Park, MD 20742}\\[.1in]
         {#3}\\[.1in] {\it {#4}}\\ {\it {#5}}\\[.4in]
         Physics Publication \#{#6}\\ {#7}\\[.5in] {\bf ABSTRACT}\\[.1in]
         \end{center} \begin{quotation}}                 
\def\border{                                            
         \setlength{\unitlength}{1mm}
         \newcount\xco
         \newcount\yco
         \xco=-21
         \yco=12
         \begin{picture}(140,0)
         \put(\xco,\yco){$\ktl$}
         \advance\yco by-1
         {\loop
         \put(\xco,\yco){$\kcr$}
         \advance\yco by-2
         \ifnum\yco>-240
         \repeat
         \put(\xco,\yco){$\kbl$}}
         \xco=158
         \yco=12
         \put(\xco,\yco){$\ktr$}
         \advance\yco by-1
         {\loop
         \put(\xco,\yco){$\kcr$}
         \advance\yco by-2
         \ifnum\yco>-240
         \repeat
         \put(\xco,\yco){$\kbr$}}
         \put(-20,13){\tiny University of Maryland Elementary Particle
Physics University of Maryland Elementary Particle Physics University of
Maryland Elementary Particle Physics}
         \put(-20,-241.5){\tiny University of Maryland Elementary
Particle Physics University of Maryland Elementary Particle Physics
University of Maryland Elementary Particle Physics}
         \end{picture}
         \par\vskip-8mm}
\def\bordero{                                           
         \setlength{\unitlength}{1mm}
         \newcount\xco
         \newcount\yco
         \xco=-31
         \yco=12
         \begin{picture}(140,0)
         \put(\xco,\yco){$\ktl$}
         \advance\yco by-1
         {\loop
         \put(\xco,\yco){$\kclr}
         \advance\yco by-2
         \ifnum\yco>-240
         \repeat
         \put(\xco,\yco){$\kbl$}}
         \xco=151
         \yco=12
         \put(\xco,\yco){$\ktr$}
         \advance\yco by-1
         {\loop
         \put(\xco,\yco){$\kcr$}
         \advance\yco by-2
         \ifnum\yco>-240
         \repeat
         \put(\xco,\yco){$\kbr$}}
         \put(-20,12){\ooo 
bacdefghidfghghdhededbihdgdfdfhhdheidhdhebaaahjhhdahba

hgdedge
    hgfdiehhgdigicba}
         \put(-20,-241.5){\ooo 
ababaighefdbfghgeahgdfgafagihdidihiidhiagfedhadbfd

ecdcdfa
    gdcbhaddhbgfchbgfdacfediacbabab}
         \end{picture}
         \par\vskip-8mm}
\def\headpic{                                           
         \indent
         \setlength{\unitlength}{.4mm}
         \thinlines
         \par
         \begin{picture}(29,16)
         \put(165,16){\line(1,0){4}}
         \put(170,16){\line(1,0){4}}
         \put(180,16){\line(1,0){4}}
         \put(175,0){\line(1,0){4}}
         \put(180,0){\line(1,0){4}}
         \put(185,0){\line(1,0){4}}
         \put(169,0){\line(0,1){16}}
         \put(170,0){\line(0,1){16}}
         \put(179,0){\line(0,1){16}}
         \put(180,0){\line(0,1){16}}
         \put(184,0){\line(0,1){16}}
         \put(185,0){\line(0,1){16}}
         \put(169,16){\oval(8,32)[bl]}
         \put(170,16){\oval(8,32)[br]}
         \put(179,0){\oval(8,32)[tl]}
         \put(185,0){\oval(8,32)[tr]}
         \end{picture}
         \par\vskip-6.5mm
         \thicklines}
\def\title#1#2#3#4{\border\headpic {\hbox to\hsize{#4 \hfill UMDEPP #3}}\par
         \begin{center} \vglue .5in {\large\bf #1}\\[.6in]
         {#2}\\[.1in] {\it Department of Physics and Astronomy}\\
         {\it University of Maryland, College Park, MD 20742}\\[1.5in]
         {\bf ABSTRACT}\\[.1in] \end{center} \begin{quotation}}  
\def\Title#1#2#3#4#5#6#7{\border\headpic
         {\hbox to\hsize{#7 \hfill UMDEPP #6}}\par
         \begin{center} \vglue .4in {\large\bf #1}\\[.4in]
         {#2}\\[.1in] {\it Department of Physics and Astronomy}\\
         {\it University of Maryland, College Park, MD 20742}\\[.1in]
         {#3}\\[.1in] {\it {#4}}\\ {\it {#5}}\\[.5in] {\bf ABSTRACT}\\[.1in]
         \end{center} \begin{quotation}}                 
\def\endtitle{\end{quotation}\newpage}                  

\def\qd{{\kern0.5pt q \kern-5.05pt \raise5.8pt\hbox{$\textstyle.$}\kern
0.5pt}}

\begin{document}

\begin{figure}[ht]
\begin{center}
\epsfig{file=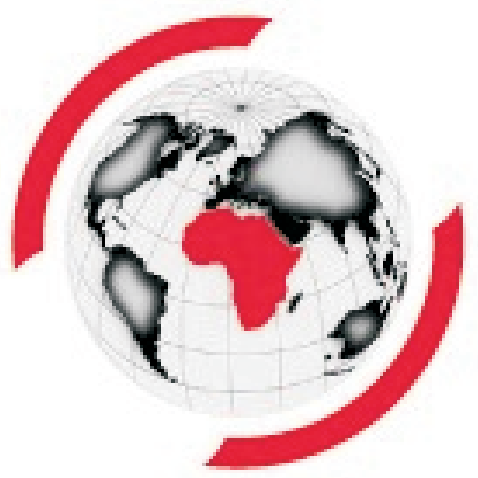,height=4cm}
\end{center}
\end{figure}

\vspace{-3cm}

\def\gfrac#1#2{\frac {\scriptstyle{#1}}
         {\mbox{\raisebox{-.6ex}{$\scriptstyle{#2}$}}}}
\def\gg{{\hbox{\sc g}}}
{\hbox to\hsize{$~$ \hfill
{STIAS-02-001}}} \par
{\hbox to\hsize{February 2002 \hfill
{UMDEPP 02-030}}} \par
{\hbox to\hsize{$~$ \hfill
{CALT-68-2366}}} \par
\setlength{\oddsidemargin}{0.3in}
\setlength{\evensidemargin}{-0.3in}
\begin{center}
\vglue .05in
{\Large\bf Is Stringy-Supersymmetry \\
Quintessentially Challenged?\footnote{
Presentation at the ``Supergravity@25'' Conference, SUNY@Stony Brook,
Dec. 1-2, 2001.}  }
\\[.25in] S.\ James Gates, Jr.\footnote{Permanent Address: Department of
Physics, University of Maryland, College Park, \\
${~~~~~~~~~~~~~~~~~~~~~~~~~~~~~~~~\,}$ MD 20742-4111 USA}${}^,$
\footnote{Supported in part by National Science Foundation Grants
PHY-01-5-23911}${}^,$ \footnote{On Sabbatical leave at the
California Inst. of Technology, Sept. 2001 thru July 2002}
${}^,$ \footnote{gatess@wam.umd.edu}
\\[0.06in]
{\it STIAS\footnote{Stellenbosch Institute of Advanced Study}
\\Private Bag XI\\
Stellenbosch, 7602 \\ Republic of South Africa}
\\[.3in]

{\bf ABSTRACT}\\[.01in]
\end{center}
~~~~We discuss the problem of introducing background spacetimes of the
de Sitter type (quintessential backgrounds) in the context of fundamental
theories involving supersymmetry.  The role of a model presented in 1984,
showing that these backgrounds can occur as spontaneously broken phases
of locally supersymmetric 4D, $N$-extended theories, is highlighted.  The
present twin challenges of the presume presence of supersymmetry in particle
physics and the emergence of experimental evidence for a positive
cosmological constant from type-II supernovae data are noted for the continued
investigation of superstring/M-theory.  Finally we note how the 1984-model
may have a role to play in future investigations.
\\[.2in]
\begin{quotation}
{}
\hspace{-1.7cm}
PACS: 03.70.+k, 11.30.Rd, 04.65.+e

\hspace{-1.7cm} Keywords: Gauge theories, Supersymmetry, Supergravity.
\endtitle

\section {Introduction}

~~~~I wish to begin by thanking the organizing committee of the
conference for the invitation to speak at this meeting.  For
twenty-five years, I have offered contributions to the work of this
community.  Just prior to the first presentation of supergravity
theory by Freedman, van Nieuwenhuizen and Ferrara \cite{A1}, as a
graduate student I turned attention to (and later published \cite{A2})
a study of supergeometry.   In those days, there was a competing
construct called ``gauge supersymmetry'' based on the notion of
extending Riemannian geometry to Salam-Strathdee superspace.  Approaching
the attainment  of my graduate degree, I had understood why the work
of Gordon Woo clearly indicated that any attempt based on Riemannian
geometry was doomed to failure.

On becoming a postdoctoral researcher as a Junior Fellow entering
Harvard's physics  department, I was already busily attempting to
understand how a Riemann-Cartan geometry might serve to by-pass
the difficulty indicated by Woo's work.  There I met Warren Siegel.
Within five minutes of meeting, we had a serious disagreement over what
must be the nature of a successful theory of curved superspace.  The
approach he was pursuing initially had no relation at all to the
notions of supergeometry.  Instead he anchored his approach on the
principles of 4D, $N$ = 1 supersymmetric gauge theory.  After getting past
our initial rough beginning, we joined forces and were able to
produce a complete theory of 4D, $N$ = 1 curved superspace \cite{A3}
that contained both the elements of supersymmetric gauge theory and
notions of supergeometry.  Warren's pre-potentials\footnote{To my
knowledge, this word was coined in one of our papers\cite{A4}.
As I recall, a referee \newline ${~~~~~}$ complained that we should at
least  explain what it meant!} were ``hiding'' inside the
geometrical entities on which I had focussed my attention.

The hosts of this meeting charged each speaker to be a bit like the
roman god Janus and make a presentation that looks forward to
unsolved problems of the future in light of past solved problems
in the field.  The story in the previous paragraph is presented
in light of our charge.  As I look at our field today and its
connection to superstring theory, I am struck at how similar
is the present state to that time before supergeometry and
pre-potentials had been successfully joined.  In 1989, I gave
a presentation the XXV Winter School of Theoretical Physics in
Karpacz, Poland.  Near the end of my contribution to the
proceedings there appears an appendix entitled, ``Treat the
String Field as a Prepotential!''  Thirteen years have passed
and we seem only a little closer to this goal in covariant string
field theory.  The story above, in my view, will ultimately be
repeated for this greatest challenge to superstring theory which
contains supergravity theory as a particular limit.  The gauge
transformations of open string theory now seem to bare a striking
resemblance to the chiral transformation ($\L$-gauge symmetries)
of superfield supergravity or superfield Yang-Mills theories.
To my mind the relation between open strings versus closed strings is
very reminiscent of that between chiral and vector superfields.
Thus, from the perspective of supergravity, the gauge parameters
of closed superstrings should be open superstrings.  So some
progress, I believe, is being made.  If my conjecture is correct,
at some point in the future there will exist a geometrical theory
of superstrings built (at least) in part on a ``string space'' possessing
stringy torsions, curvatures and field strengths.  These are likely
to possess constraints whose solutions will, indeed, have the string
field as their prepotential solution.  In others words, I believe
superstring theory will ultimately follow the supergravity paradigm.

On the occasion of the first day of this meeting, our community has
drawn itself together.  In a sense the various members of this
community seem almost like members of a single family.  (We even
have had some terrific fights in the past to prove it.)  But still
the community has struggled to make its contributions to the
advancement of our field and is in a healthy state today with still
unmet challenges ahead.

\section {Non-supersymmetric Preliminary Remarks}

~~~~The cosmological constant $\l$ has provided a topic of lively
debate almost continuously since its introduction into the physics
literature \cite{R1}.  It enters the Einstein Field Equations\footnote{We
use the notations and  conventions of {\it {Superspace}} (e.g. the
gravitational coupling constant
\newline ${~~~~~~}$ $\k^2 = 48 \pi G$ \cite{R2}).} as
\be  \eqalign{
{\cal R}_{\un a \, \un b} ~-~ \fracm 12 \eta_{\un a \un b}
\, {\cal R} ~+~ \l \eta_{\un a \un b} ~=~ - \, \fracm 16 \,
\k^2 \, {\cal T}_{\un a \, \un b}
  ~~~. } \label{eq:01} \ee
At the time of its introduction, Einstein stated, ``That term is necessary
only for the purpose of making possible a quasi-static distribution of
matter, as required by the fact of the small velocities of the stars.''
Of course, this reason has long since disappeared.  With the discovery of
the expansion of the universe, Einstein later described the cosmological
constant as, ``the biggest blunder he ever made in his life.''

However, within the present epoch of particle physics there has been a
``cosmological constant problem.''  To most simply see this problem it
suffices to use a toy model that is easy to construct by considering
a self-interacting spin-0 field ($\phi$) in the presence of gravitational
interactions. We may write a model of the form ${\cal S}_{Tot} = {\cal
S}_{grav} + {\cal S}_{matter}$ where
\be  \eqalign{
{\cal S}_{grav} &=~ - \fracm 3{\k^2} \, \int d^4 x \, {\rm e}^{-1} ~
  \Big[ ~ {\cal R}(e) ~-~ \l ~ \Big] ~~~, \cr
{\cal S}_{matter} &=~ \int d^4 x \, {\rm e}^{-1} ~ \Big[ \,- \fracm 14
(e^{\un a} \phi\,) (e_{\un a} \phi\,) ~-~ (\, \fracm 1{4!} \l_0 \phi^4
\, + \, \fracm 12 m_0^2 \phi^2 \,) \,~ \Big]
  ~~~. } \label{eq:02} \ee
Upon examination of the equation of motion for $\phi$, with the additional
assumption that $\phi$ has a nonvanishing vacuum expectation value $\VEV
\phi $, we find that $V' (\VEV \phi ) = 0$ (with $ V(\phi) = \fracm 1{4!}
\l_0 \phi^4  \, + \, \fracm 12 m_0^2 \phi^2$) defines the ground state.
If $m_0^2 >0$ then $\VEV \phi  = 0$ and the cosmological constant that
appears is the second term in the first action describing the ``vacuum''
spacetime (a space of constant curvature completely described by $\l$).
This vacuum spacetime is flat when $\l = 0$.  On the other hand, things
change markedly if $m_0^2 <0$ (as in the case of spontaneous symmetry
breakdown).  In this case, $\VEV \phi  \ne 0$ and the vacuum spacetime
has a cosmological constant given by $\l_{tot} = - \l + \frac{\k^2}{3} V
(\VEV \phi )$.  Evaluated at the vev of $\phi$, the potential takes the
form $V (\VEV \phi ) = - (3/2) \l_0 (m_0^4/\l_0^2)$ so $\l_{tot} = \l
[  1 + \l_0 (\k^2  m_0^4/2 \l \l_0^2)]$.   Since $\l$, $\l_0$ and $m_0$ are
completely free parameters, $\l_{tot}$ can describe a de Sitter space
($\l_{tot}$ $>$ 0), anti-de Sitter  space ($\l_{tot}$ $<$ 0) or Minkowski
space ($\l_{tot} = $ 0).  We are thus able to ``adjust''  or ``tune''
the parameters of effective cosmological constant $\l_{tot}$.

At the level of non-quantum classical considerations, the tuning of
the original cosmological constant to achieve a vacuum spacetime that
is flat may be considered a matter of taste.  At the quantum level,
there is the technical matter of considering the renormalized values
of the bare parameters that appear above.  In particular in the presence
of quantum corrections, it is natural to expect a renormalized equation
of the form $\l_{tot} = c_1 \l + c_2 \frac{\k^2}{3} V (\VEV \phi )$
where $c_1$ and $c_2$ are constants determined by quantum corrections
to the theory.  There is no reason to expect the continued equality
of $c_1$ and $c_2$ and so to define  a flat vacuum spacetime would require
``re-tuning'' (i.e. using a  value for $\l$ that is different from
that  in the non-quantum theory).

It is true that within ordinary gravity, the non-renormalizability of
the theory intrudes into this argument.  However, with the proposal
of `eka-general relativity,' such as superstring theory (or some as
yet unknown construction), we might be forced to squarely face this
problem still.

The value of the cosmological constant also shows up in a fundamental way
in the structure of the spacetime symmetries of the universe.  In a
theory of gravitation\footnote{For additional discussion of the gauge
approach to gravitational theories see \cite{R3}}, a gauge covariant
derivative
\be  \eqalign{
\nabla_{\un a} ~ \equiv~ {\rm e}_{\un a}{}^{\un m} \pa_{\un m}
~+~ \o_{\un a ~ \g }{\,}^{\d } \,{\cal M}_{\d }{}^{\g} ~+~
\o_{\un a ~ \Dot \g }{\,}^{\Dot \d} \, {\Bar {\cal M}}_{\Dot
\d}{}^{\Dot \g}
  ~~~,} \label{eq:03} \ee
can be introduced.  In a spacetime with vanishing torsion (thus
determining the spin-connections, $\o_{\un a \g \d } $ and $\o_{\un a
\Dot \g \Dot \d} $ in term of the vierbein $ {\rm e}_{\un a}{}^{\un
m}$), the commutator algebra of this derivative takes the form
\be  \eqalign{
\Big[ ~ \nabla_{\un a} \, , \, \nabla_{\un b} ~ \Big]
&=~ {\cal R}_{\un a \, \un b ~ \g}{\,}^{\d}\, {\cal M}_{\d}{}^{\g}
~+~ {\cal R}_{\un a \, \un b ~ \Dot \d}{\,}^{\Dot \g} \,\, {\Bar
{\cal M}}_{\Dot \d}{}^{\Dot \g}
  ~~~, } \label{eq:04} \ee
and due to the reality of the derivative, the field strength ${\cal
R}_{\un a \, \un b \, \Dot \d \, \Dot \g} $ is just the complex
conjugate of ${\cal R}_{\un a \, \un b \, \g \, \d}$.

The vacuum configuration of the gauge-gravitational covariant
derivative in (\ref{eq:03}) is a specification of the vierbein
field.  One class of such field configurations is described by,
\be  \eqalign{
\nabla_{\un a} ~=~ \Big[ ~ 1 \,-\, \fracm 16 \l \, x^2 ~ \Big] \,
\pa_{\un a} ~-~ \fracm 13 \l \, x_{\g \Dot \a} \,  {\cal M}_{\a}
{}^{\g}  ~-~  \fracm 13 \l  \, x_{\a \Dot \g} \, {\Bar {\cal
M}}_{\Dot \a}{}^{\Dot \g}
  ~~~, } \label{eq:05} \ee
dependent on the cosmological constant of the dimensions of
mass-squared. Comparing (\ref{eq:03}) with (\ref{eq:05}), the
spin-connection terms proportional to ${\cal M}$ and ${\Bar {\cal
M}}$ in the latter are chosen so that the torsion term vanishes
consistently with (\ref{eq:04}).

When this field configuration is substituted into
(\ref{eq:04}), we find
\be  \eqalign{
\Big[ ~ \nabla_{\un a} \, , \, \nabla_{\un b} ~ \Big] &=~  -
  \fracm 23 \l  \, \Big[ ~ C_{{\Dot \a} \ {\Dot \b}} \, {\cal M}_{\a
\, \b} ~+~ C_{\a \, \b} \, {\Bar {\cal  M}}_{{\Dot \a} \,
{\Dot \b}} ~ \Big]  ~~~, } \label{eq:06} \ee
and thus the configuration in (\ref{eq:05}) describes a spacetime
in which the Riemann curvature tensor is a constant
\be  \eqalign{
{\cal R}_{\un a \, \un b ~ \g}{\,}^{\d} ~=~   \fracm 13 \l  \,
C_{{\Dot \a} \ {\Dot \b}} ~ \Big[ \,  C_{\a \g} \, \d_{\b}{}^{\d}
~+~  C_{\b \g} \, \d_{\a}{}^{\d}  \, \Big]
  ~~~. } \label{eq:07} \ee
Finally, the field configuration in (\ref{eq:05}) may be inserted
into the Einstein Field Equation (\ref{eq:01}).  The equation is
found to be satisfied if
\be  \eqalign{
  {\cal T}_{\un a \, \un b}  ~=~ 0
  ~~~. } \label{eq:08} \ee

A relation to the symmetries of spacetime comes about as follows.
The gauge-gravitational covariant derivative may be used to define
``covariant translation'' operators ${\cal P}{}_{\un a}$ = $i \nabla_{
\un a}$.   Accordingly, the commutator algebra of the translation
operators is fixed according to (\ref{eq:03} - \ref{eq:06}) so that
\be  \eqalign{
\Big[ ~ {\cal P}{}_{\un a} \, , \, {\cal P}{}_{\un b} ~ \Big] &=~
  \fracm 23 \l  \, \Big[ \, C_{{\Dot \a} \ {\Dot \b}} \, {\cal M}_{\a
\, \b} ~+~ C_{\a \, \b} \, {\Bar {\cal  M}}_{{\Dot \a} \,
{\Dot \b}} ~ \Big]  ~~~. } \label{eq:09} \ee
In this way the translational symmetry of spacetime is sensitive
to the value of the cosmological constant.  For $\l$ = 0 the case
of Minkowski space, the translation generators form an abelian
group.  For $\l $ $>$ 0 de Sitter or $\l$ $<$ 0 anti-de Sitter
spaces, the translation generators together with  the spin-angular
momentum generators ${\cal M}_{\a \b}$ and ${\Bar {\cal M}}_{{\Dot
\a}{\Dot \b}}$ form non-abelian groups.

Finally, there is one other kinematical feature that is of note
in the issue of de Sitter spacetimes and anti-de Sitter spacetimes
vis-a-vis their relation to Minkowski spacetimes.  Massless
representations in the Minkowski space of the Poincar\' e group with
generators $P_{\un a}$, $ J_{\a \b}$ and $ {\Bar J}{}_{{\Dot \a}
{\Dot \b}}$ and commutation algebra\footnote{A set of super-vector
fields that provide a representation for the generators of the
superconformal \newline ${~~~~\,}$ group can be found on pages
76 and 81 of \cite{R2}.},
\be  \eqalign{
\Big[ ~ P_{\un a} \, , \, P_{\un b} ~ \Big] &=~ 0 ~~~,~~~ \Big[
~ J_{\a}{}_{\b} \, , \, J_{\g}{}_{ \d} ~ \Big] ~=~  i \, [ \,
C_{\b \g} \, J_{\a}{}_{ \d} ~+~ C_{\a \, \d} \, J_{\b}{}_{ \g}
  \, ] ~~~, \cr
\Big[ ~ J_{\a \, \b} \, , \, {\Bar J}{}^{\Dot \g \Dot \d} ~
\Big] &=~ 0 ~~~,~~~ \Big[ ~ {\Bar J}{}_{\Dot \a}{}_{\Dot \b}
\, , \, {\Bar J}{}_{\Dot \g}{}_{\Dot \d} ~ \Big] ~=~   i \, [ \,
C_{\Dot \b \, \Dot \g} \, {\Bar J}_{\Dot \a}{}_{ \Dot \d} ~+~
C_{\Dot \a}{}_{\Dot \d}  \,\,  {\Bar J}_{\Dot \g}{}_{
\Dot \b} \, \, ] ~~~, \cr
\Big[ ~ J_{\a \, \b} \, , \, P_{\un c} ~ \Big] &=~ i \fracm 12
C_{\g (\a} P{}_{\b) \Dot \g} ~~~,~~~ \Big[ ~ {\Bar J}{}_{\Dot \a
\, \Dot \b} \, , \, P_{\un c} ~ \Big] ~=~ i \fracm 12
C_{\Dot \g ( \Dot \a |} P{}_{\g | \Dot \b) } ~~~, }
\label{eq:10} \ee
also form representation of the larger conformal group with additional
generators $K_{\un a}$ and $\D$ and the enlarged commutator algebra
\be  \eqalign{  {~~~~~~~~}
\Big[ ~ \D \, , \, P_{\un b} ~ \Big] &=~ i \, P{}_{\un c}  ~~~,~~~
\Big[ ~ \D \, , \, K_{\un b} ~ \Big] ~=~ - i \, K{}_{\un c}
~~~, ~~~ \Big[ ~ K{}_{\un a} \, , \, K{}_{\un b} ~ \Big] ~=~ 0 ~~~,\cr
\Big[ ~ J_{\a \, \b} \, , \, K{}_{\un c} ~ \Big] &=~ i \fracm 12 \,
C_{\g (\a} K{}_{\b) \Dot \g} ~~~,~~~ \Big[ ~ {\Bar J}{}_{\Dot \a \,
\Dot \b} \, , \, K{}_{\un c} ~ \Big] ~=~ i \fracm 12 \,
C_{\Dot \g ( \Dot \a | } K{}_{\g | \Dot \b) } ~~~,  \cr
\Big[ ~ P{}_{\un a} \, , \, K{}_{\un b} ~ \Big] &=~ i
\, (\, C_{\Dot \a}{}_{\Dot \b} \, J_{\a}{}_{ \b} ~+~
C_{\a}{}_{ \b}  \, J_{\Dot \a}{}_{\Dot \b}
~+~ \eta_{\un a}{}_{\un b}  \,  \D  \, ) ~~~, \cr
\Big[ ~\D \, , \, J_{\a \, \b}  ~ \Big] &=~ 0 ~~~, ~~~
\Big[ ~\D \, , \, {\Bar J}_{\Dot \a}{}_{\Dot \b}  ~ \Big] ~=~ 0
  ~~~. }
\label{eq:11}
\ee

The point to note is that the generators of translations in both
anti-de Sitter and de Sitter spacetimes denoted by ${\cal P}_{\un a}$
may be regarded as a linear combination of the generators $P_{\un
a}$ and $K_{\un a}$ in the Minkowski spacetime,
\be  \eqalign{
{\cal P}_{\un a} ~\equiv~ P_{\un a} ~\pm~ \fracm 13 | \l | \, K_{\un a}
  ~~~, }   \label{eq:12} \ee
where a parameter with the dimensions of $\l$ {\it {must}} be
introduced owing to the difference in dimensions of $P_{\un a}$
and $K_{\un a}$. It is a directly simple calculation to begin with
the definition  in (\ref{eq:12}) and the commutator algebra in
(\ref{eq:10}) and (\ref{eq:11}) to thusly prove
\be  \eqalign{
\Big[ ~ {\cal P}{}_{\un a} \, , \, {\cal P}{}_{\un b} ~ \Big] &=~
\pm \, i \, \fracm 23 | \l | \, \Big[ \, C_{{\Dot \a} \ {\Dot \b}}
\, J{}_{\a \, \b} ~+~ C_{\a \, \b} \, {\Bar J}{}_{{\Dot \a} \,
{\Dot \b}} ~ \Big]  ~~~, }   \label{eq:13}  \ee
and the sign of the linear combination in (\ref{eq:12}) is seen
to determine whether the translation generator ${\cal P}_{\un a}$ is
related to an anti-de Sitter or de Sitter geometry.

The discussion above might have been deemed solely formal and of little
importance except that the recent experimental data from type-II super
novae \cite{R4,R5} seem to indicate that we live in a universe that
possesses a small {\it {positive}} cosmological constant (de Sitter
geometry).  Thus, the configuration in (\ref{eq:05}) apparently
describes our universe in the limit of no gravitational radiation.

\section {Joining the Clash: de Sitter vs. SUSY}

~~~~Soon after the introduction of supergravity theories, Ferrara
\cite{R6} noted an interesting distinction that global supersymmetry
makes with regard to spacetimes of constant curvature.  Namely global
supersymmetry can easily be realized for anti-de Sitter or Minkowski
spaces, but cannot be realized at all for de Sitter spaces.   Although,
Ferrara cast his discussion in  terms of charges and supercharges,
we can understand the gist of his discussion by probing the structure
of superspace supergravity  covariant derivatives that are consistent
with a superspaces that contains a bosonic spacetime of constant
curvature.

The superanalog of (\ref{eq:03}) takes the form of
\be  \eqalign{
\nabla_{\un A} ~ \equiv~ {\rm E}_{\un A}{}^{\un M}
\pa_{\un M} ~+~ \o_{\un A ~ \g }{\,}^{\d } \,{\cal M}_{\d }{}^{\g
}  ~+~ \o_{\un A ~ \Dot \g }{\,}^{\Dot \d} ~ {\Bar {\cal
M}}_{\Dot \d}{}^{\Dot \g}
  ~~~,} \label{eq:14} \ee
where the super-index $\un A$ is permitted to take on values $\a$,
$\Dot \a$ and $\un a$ and the results in \cite{R6} are equivalent
to the statement that the most general super-commutator algebra
that is consistent with a bosonic subspace of constant curvature
must take the form,
\be  \eqalign{
[ \, \nabla_{\a} ~,~ \nabla_{\b} \,  \} &=~ - \, 2 \, \ell_{\cal F}
\, {\cal M}_{\a \b} ~~~,~~~ \cr
[ \, \nabla_{\a} ~,~ {\Bar \nabla}{}_{\Dot
\a} \,  \} &=~ i \nabla_{\un a} ~~~, \cr
[ \, \nabla_{\a} ~,~ \nabla_{\un b} \,  \} &=~ - \, \ell_{\cal F} \,
C_{\a \b} \, {\Bar \nabla}{}_{\Dot \b} ~~~, \cr
[ \, \nabla_{\un a} ~,~ \nabla_{\un b} \,  \} &=~ 2 \, |
\ell_{\cal F} |^2 \, ( ~C_{\Dot \a \Dot \b} \, {\cal M
}_{\a \b} ~+~ C_{\a \b} \,  {\Bar {\cal M }}{}_{\Dot \a \Dot \b}
~)  ~~~.} \label{eq:15} \ee
where $\ell_{\cal F}$ is a constant parameter (the ``Ferrara parameter'').
Upon comparing the last line here with the result in (\ref{eq:06}), we
see that the relation  between the cosmological constant and the Ferrara
parameter is\footnote{Up to a complex phase, Ferrara's parameter is the
square root of the cosmological constant.}
\be  \eqalign{
- \l  ~=~ 3 \,| \ell_{\cal F} |^2
  ~~~.} \label{eq:16} \ee
Since the rhs of this equation is non-negative, the equation only has
non-trivial solutions if $\l$ $\le$ 0, i.\ e.\ flat spaces or anti-de Sitter
spaces. The consistency of these is verified by checking the super Bianchi
identities for the graded commutators in (\ref{eq:15}).  Upon comparison
between this result and the one in (\ref{eq:07}) we see that this superspace
result implies that the bosonic subspace contained within it must be either a
Minkowski space ($\ell_{\cal F}$ = 0) or a space of constant curvature
($\ell_{\cal F}$ $\neq$ 0) with the Ricci tensor and curvature scalar
respectively taking the forms ${\cal R}_{\un a \un b}$ = $-|\l| g_{\un
a \un b}$ and  ${\cal R}$ = $-4 |\l|$.

It is easily realizable that for no value of the Ferrara parameter is
it possible to obtain a cosmological constant of an appropriate sign
so as to describe a de Sitter geometry.  We should perhaps mention
that this is a robust model-independent result which in its original
presentation was cast in the language of charges and group theory.
It essentially forms a  no-go theorem for the realization of rigid
supersymmetry in the presence of a de Sitter spacetime.

One of the hallmarks of supersymmetry is that the usual translation
operator $P{}_{\un a}$ is related to spinorial supercharges ${
Q}_{\a}$ and ${\Bar Q}_{\Dot \a}$ via the equation,
\be  \eqalign{
[~ Q{}_{\a}\, , \,{\Bar Q}_{\Dot \a} ~ \} ~=~ P{}_{\un a}
  ~~~.} \label{eq:17} \ee
In the context of superconformal supersymmetry there is a similar
relation between $K{}_{\un a} $ and the s-supersymmetry generator
\be  \eqalign{
[~ S{}_{\a}\, , \,{\Bar S}_{\Dot \a} ~ \} ~=~ K{}_{\un a}
  ~~~.} \label{eq:18} \ee
The other relevant graded commutator takes the form
\be  \eqalign{
[~ Q{}_{\a}\, , \, S_{\b} ~ \} ~=~ - i \, (\, J{}_{\a}{}_{\b}
~+~ \fracm 12 \, C_{\a}{}_{\b} \, \D  \,) ~-~ \fracm 12 \,
C_{\a}{}_{\b} \, {\cal Y}
  ~~~.} \label{eq:19} \ee

In a space of constant curvature, there is a supercharge ${\cal Q}
{}_{\a}$ satisfying
\be  \eqalign{
[~ {\cal Q}_{\a}\, , \,{\Bar {\cal Q}}{}_{\Dot \a} ~ \} ~=~ {\cal
P}_{\un a}
  ~~~,} \label{eq:20} \ee
and it can be constructed from the superconformal generators
$Q{}_{\a}$ and $S{}_{\a}$ via
\be  \eqalign{
{\cal Q}_{\a} ~=~ Q{}_{\a}  ~+~  \ell_{\cal F} \, S{}_{\a}
~~~,} \label{eq:21} \ee
in analogy with the construction in (\ref{eq:12}).  Here the Ferrara
parameter is seen to determine by how much is the anti-de Sitter
supersymmetry generator ${\cal Q}_{\a}$ ``deformed''  from
the supersymmetry generator $Q{}_{\a}$ by the
s-supersymmetry generator $S{}_{\a}$.  Using this expression to calculate
$ {\cal P}_{\un a}$ (together with the fact that $[Q, {\Bar S} \} = 0$)
in (\ref{eq:20}) yields
\be  \eqalign{
{\cal P}_{\un a} ~\equiv~ P_{\un a} ~+~ | \ell_{\cal F} |^2 \, K_{
\un a}  ~\to~  {\cal P}_{\un a} ~\equiv~ P_{\un a} ~-~ \fracm 13 \l
\, K_{\un a}   ~~~.
} \label{eq:22} \ee
The representation theory of superspace, as encoded in the set of
supervector fields \cite{R2} that realize the generators, necessarily
has picked anti-de Sitter space as the type of constant curvature space
consistent with supersymmetry.  This is true independent of the phase
of the Ferrara parameter since only its absolute modulus enters
(\ref{eq:22}) above.   In comparing (\ref{eq:12}) to (\ref{eq:22}),
here only the linear combination corresponding to anti-de Sitter
backgrounds occurs and this is solely due to supersymmetry.

\section {The Era of No De Sitter Space for Local \\
${}$N-extended SUSY}

~~~~Since the organizers of this conference charged the presenters with
the dual tasks of looking back at some supergravity history in order
to address current and possibly future developments, this section will
be spent giving a review of the topic of spaces of constant curvature
in the context of supergravity theory.  The 4D, $N$ = 4 supergravity
theory played a special role in this line of investigation so we will
also review its development.

During early explorations on the issue of the cosmological constant in
supergravity theory, there was ample support for Ferrara's observation.
Shortly after supergravity appeared, Freedman \cite{R7} presented
the first discussion of the theory's consistency in a space with a
non-vanishing cosmological constant.  He found that only anti-de Sitter spaces
were admitted if R-symmetry is also gauged.  A similar construction
by Townsend \cite{R8}, without the gauging of R-symmetry, reached the
same conclusion.  Support for the absence of de Sitter space in local
supersymmetrical theories continued to be found in numbers of other
works during these early days.  The gauging of the SO(2) and SO(3)
automorphism groups of 4D, $N$ = 2 and $N$ = 3 supergravities,
respectively, by Freedman and Das \cite{R9} also ruled out de Sitter
backgrounds.

The theory of 4D, $N$ = 4 supergravity first appeared in its ungauged
version, but it was immediately recognized that there were new aspects
of the theory.  First a version admitting an SO(4) automorphism group
was found \cite{R10} and then a second version with an SU(2) $\otimes$
SU(2) automorphism appeared \cite{R11}.  Much later \cite{R12}, even
a third version of the ungauged theory was found.  In view of our later
discussion of investigations into some unresolved issues, we believe
for clarity's sake a discussion of the relations between these three
{\it {ungauged}} automorphism versions of 4D, $N$ = 4 theories
is warranted here.

Cremmer, Scherk and Ferrara \cite{R11} showed a relation between the
first two discovered versions of the theory which are connected via
a conformal map.  There are two spin-0 fields in both of these theories.
Let us denote the scalar and psuedoscalar fields in the version with
the SO(4) automorphism group by $A^{\prime}$ and $B^{\prime}$.  Similarly
we denote the scalar and psuedoscalar fields in the version with the
SU(2) $\otimes$ SU(2) automorphism group by $A$ and $B$.  Let two
complex variables  $\cal W$ and $\cal Z$ be defined by
\be  \eqalign{
&{~~~} {\cal W} ~\equiv~ A(x) ~+~ i B(x) ~~~, ~~~
{\cal Z} ~\equiv~ A^{\prime}(x) ~+~ i B^{\prime}(x) ~~~,
} \label{eq:23} \ee
and under a conformal mapping defined by
\be  \eqalign{
{\cal Z} ~\to~  { {\cal W} \over {{\cal W} \, - \, 1 }} ~~~,
} \label{eq:24} \ee
induces the following change on the spin-0 kinetic energy terms,
\be  \eqalign{
&{~~~}
{ 1 \over { 1 \, - \, | {\cal Z} |^2 }} |\pa {\cal Z} |^ 2
~~~\to ~~~
{ 1 \over {\sqrt { 1 \, - \, {\cal W}  \, - \,  {\Bar {\cal W}
}  }}  }   | \pa {\cal W} |^ 2 ~~~.
} \label{eq:25} \ee

If we perform the same conformal map on the last part of (\ref{eq:24})
\be  \eqalign{
{\cal W} ~\to~  { {\cal Z} \over {{\cal Z} \, - \, 1 }} ~~~,
} \label{eq:26} \ee
it will then be transformed into the first part.  This conformal map
is the same as in (\ref{eq:24}) and applying this to (\ref{eq:24})
results in the identity transformation on $\cal Z$.  This is an example
of an idempotent conformal mapping.  Apply this map and some related ones
acting on other fields in the SU(2) $\otimes$ SU(2) action then
transform it into the SO(4) action and vice-versa,
\be  \eqalign{
{\cal S}_{ {\rm {SU(2)}} \otimes {\rm {SU(2)}}}({\cal W}, \, ...)
~~\leftrightarrow ~~ {\cal S}_{ {\rm {SO(4)}} }({\cal Z}, \, ...)
  ~~~,
} \label{eq:27} \ee
where the ellipsis denote all the remaining fields in each action.

Within the confines of the SU(2) $\otimes$ SU(2) version, the field $B$
{\it {only}} appears in the Lagrangian and transformation laws according
to the form $\pa_{\un a} B$.  Due to this, Nicolai and Townsend \cite{R12} 
were able to show the existence of a third formulation which is connected
to the SU(2) $\otimes$ SU(2) formulation via an idempotent ``Hodge duality''
map.  The essence of the hodge duality map can been seen by looking at
a toy example.

Consider the two action ${\cal S}_0$ and ${\cal S}_1$ respectively defined
by
\be  \eqalign{
&~~ {\cal S}_0 ~=~ -  \, \fracm 12 \, \int d^4 x ~  f^{\un a}
\, f_{\un a} ~~~, ~~~ f_{\un a } ~\equiv~ \pa_{\un a} B ~~~,  \cr
&~~ {\cal S}_2 ~=~  \fracm 12 \, \int d^4 x ~ h^{\un a} \, h_{\un a}
~~~,~~~h_{\un a } ~\equiv~ \fracm 1{3!} \e_{\un a} {}^{\un b \,
\un c \, \un d \, } h_{\un b \, \un c \, \un d \, } ~~~, \cr
&~~~~~~~~~~~\, h_{\un b \, \un c \, \un d \, } ~=~ \pa_{\un
a} B_{\un b \, \un c} ~+~ \pa_{\un b} B_{\un c \, \un a} ~+~ \pa_{\un
c} B_{\un a \, \un b} ~~~.
} \label{eq:28} \ee
The equations of motion that follow from the extremization of these
actions are;
\be  \eqalign{
& \parvar{S_0}{B} ~=~ 0 ~~ \to ~~  \pa^{\un a} f_{\un a}
~=~ 0 ~~~, \cr
& \parvar{S_2}{B_{\un a \, \un b}} ~=~ 0 ~~ \to ~~
\pa_{\un a} h_{\un b}  ~-~ \pa_{\un b}
h_{\un a} ~=~ 0 ~~~,
} \label{eq:29} \ee
where we have expressed the equations of motion in terms of the ``field
strengths'' $f_{\un a}$ and $h_{\un a}$.  Independent of the dynamics,
these field strengths satisfy differential equations that may be called
``Bianchi identities''
\be  \eqalign{
0 & =~ \pa_{\un a} f_{\un b}  ~-~ \pa_{\un b}
f_{\un a} ~~~, \cr
0 & =~ \pa^{\un a} h_{\un a} ~~~.
} \label{eq:30} \ee
A simple comparison between (\ref{eq:29}) and (\ref{eq:30}) shows that
the equations of motion and Bianchi identities are exchanged under the
idempotent mapping
\be  \eqalign{
f_{\un a} ~~ \to ~~  i h_{\un a} ~~~,~~~ h_{\un a} ~~ \to ~~  - i f_{\un a}
~~~. } \label{eq:31} \ee

The relation between the SU(2) $\otimes$ SU(2) version and the
Nicolai-Townsend version of 4D, $N$ = 4 supergravity is analogous
to the relation between ${\cal S}_0$ and ${\cal S}_2$ above.
Under the action of the transformation in (\ref{eq:31}) and some
related re-definitions on other fields
\be  \eqalign{
{\cal S}_{ {\rm {SU(2)}} \otimes {\rm {SU(2)}}}(B, \, ...)
~~\leftrightarrow ~~ {\cal S}_{ N-T }(B_{\un a \, \un b}, \, ...)
  ~~~,
} \label{eq:32} \ee
where again the ellipsis denote all the remaining fields in each action.

Finally, the gauging of the automorphisms of the SO(4) version was carried
out by Das, Fischler and Ro\v cek \cite{R13} and for the SU(2) $\otimes$
SU(2) version by Freedman and Schwarz \cite{R14}.  In the former case, a
single non-Abelian coupling constant $g$ is required.  In the latter case,
two non-Abelian coupling constants $g_1$ and $g_2$ are required.   In both
of these cases it was found that  a potential involving the spin-0 zero
field is induced relative to the ungauged action.  In the limit where the
coupling constants vanish, these potentials go to zero.  Finally, in
neither case were de Sitter backgrounds found to satisfy the full set
of equations of motion.

The gauging of the automorphism also plays one other important role.
The three version of 4D, $N$ = 4 supergravity discussed above are all
connected to each other by a set of field re-definitions.  So naively,
one might think that they are all equivalent.  This is true as long as
the automorphism groups are ungauged.  The gauging of the automorphism
group destroys this equivalence.

For none of the models discussed above, was it possible to show that
de Sitter space occurred as a solution to the resultant equations
of motion.   Had Ferrara's Theorem been the last word on this topic,
the observation of a quintessential cosmological constant would, in
and of itself, rule out the relevance of supergravity (and hence any
theory containing it) to a description of Nature.

\section {Breakthrough: De Sitter Space As Spontaneously
Broken Phase of Local SUSY}

~~~~In 1982, we began to study the 4D, $N$ = 4 supergeometry and
discovered that from this perspective all component level formulation
were derivable within a universal setting.  Our study was motivated by
the fact that with the plethora of versions known, it was possible
that even more (then unknown) versions might also exist.  For example,
the Cremmer-Scherk-Ferrara conformal map is only one example of an
idempotent map.  The most general member has the form
\be  \eqalign{
{\cal Z}^{\prime} ~\to~  { {{\cal Z} \,+\, e^{ \varphi_0}}\over
{ e^{- \varphi_0} \, {\cal Z} \, - \,
1 }} ~~~, } \label{eq:33} \ee
dependent on the complex parameter $\varphi_0$.  Does this imply the
existence of an entire family of models?

To answer this question, we relied upon first a careful supergeometrical
analysis \cite{R15} of the then known theories\footnote{Modifications to
this analysis was provided later \cite{R16}.}.  After this analysis was
begun, a classification scheme presented itself and it became clear that
there were indeed some overlooked $N$ = 4 models.  Some of these had the
distasteful feature of possessing spin-0 kinetic energy terms of the
form
\be  \eqalign{
&{~~~} {\cal L}_{kin.} ~=~
{ 1 \over {  | {\cal Z} |^2  \, - \, 1 }} |\pa {\cal Z} |^ 2
~~~,
} \label{eq:34} \ee
and were thus not commented upon\footnote{However, these models were
subsequently discussed in the literature \cite{R17}}.  However, one of
the overlooked models \cite{R18} possessed a feature that had never
been seen for extended supergravity...a de Sitter background emerged
as a solution to the equations of motion in the presence of the
spontaneous breaking of all four local supersymmetries.  In the
following, a discussion of the technique used  for this discovery is
described.

To provide a description of 4D, $N$ = 4 supergravity, it is first
necessary to introduce the appropriate generalization of the superspace
derivative in (\ref{eq:14}).  This is done with a superspace
supergravity covariant derivative of the form,
\be  \eqalign{
&\nabla_{\a i} ~=~ {\rm E}{}_{\a i}{}^{\underline M} D_{\un M} ~+~
\o_{\a i \, \g}{}^{\d} {\cal M}_{\d}{}^{\g}~+~ \o_{\a i \, \Dot
\g}{}^{\Dot \d} {\Bar {\cal M}}_{\Dot \d}{}^{\Dot \g} ~+~ \G_{\a
\, i}{}^{ k l} \, {\cal T}_{k l} ~~~, \cr
&\, \,\nabla_{ \un a } ~=~ {\rm E}{}_{ \un a }{}^{\underline M} D_{\un
M} ~+~ \o_{\un a \, \g}{}^{\d} {\cal M}_{\d}{}^{\g}~+~ \o_{\un a
\, \Dot \g}{}^{\Dot \d} {\Bar {\cal M}}_{\Dot \d}{}^{\Dot \g}
~+~ \G_{\un a \, i}{}^{ k l} \, {\cal T}_{k l}
~~~. } \label{eq:35} \ee
Graded commutation of these operators produce the usual
torsion, curvature and field strength supertensors.
\be  \eqalign{
&[ \, \nabla_{ \a i} ~,~ {\nabla}{}_{\b j} \,  \} ~=~ T_{\a i \,
\b j}{\,}^{\d l}\, \nabla_{\d l} ~+~ T_{\a i \, \b j}{\,}^{
\un d}\, \nabla_{\un d} ~+~ {\cal R}_{\a i \, \b j ~ \g}{\,}^{\d}\,
{\cal M}_{\d}{}^{\g} \cr
&{~~~~~~~~~~~~~~~~~~~~~}~+~ {\cal R}_{\a i \, \b j ~ \Dot \d}{\,
}^{\Dot \g} \,\, {\Bar {\cal M}}_{\Dot \d}{}^{\Dot \g} ~+~ C_{\a
\b} \, {\cal E}_{i j}{}^{k l} \, {\cal  T}_{k l} ~~~, \cr
&\, [ \, \nabla_{ \a i} ~,~ {\nabla}{}_{\un b} \,  \} ~=~ T_{\a i \,
\un b }{\,}^{\d l}\, \nabla_{\d l} ~+~ T_{\a i \, \un b }{\,}^{
\un d}\, \nabla_{\un d} ~+~ {\cal R}_{\a i \, \un b ~ \g}{\,}^{\d}\,
{\cal M}_{\d}{}^{\g} \cr
&{~~~~~~~~~~~~~~~~~~~~~}~+~ {\cal R}_{\a i \, \un b  ~ \Dot \d}{\,
}^{\Dot \g} \,\, {\Bar {\cal M}}_{\Dot \d}{}^{\Dot \g} ~+~ ~+~ F_{
\a i \,  \un b}{}^{k l} \,{\cal  T}_{k l} ~~~, \cr
&~ [ \, \nabla_{ \un a } ~,~ {\nabla}{}_{\un b } \,  \} ~=~ T_{\un a
\, \un b }{\,}^{\d l}\, \nabla_{\d l} ~+~ T_{\un a  \, \un b }{\,}^{
\un d}\, \nabla_{\un d} ~+~ {\cal R}_{\un a  \, \un b  ~ \g}{\,}^{\d}\,
{\cal M}_{\d}{}^{\g} \cr
&{~~~~~~~~~~~~~~~~~~~~~}~+~ {\cal R}_{\un a  \, \un b  ~ \Dot \d}{\,
}^{\Dot \g} \,\, {\Bar {\cal M}}_{\Dot \d}{}^{\Dot \g} ~+~ F_{
\un a \, \un b}{}^{k l} \, {\cal  T}_{k l} ~~~. \cr
} \label{eq:36} \ee
In comparison to the superspace supergravity covariant derivative
of simple supergravity (\ref{eq:14}) a major difference here is
the presence of a super 1-form connection ($ \G_{\un A \, i}{}^{
k l}$) and its corresponding super 2-form field strength ($
F_{\un A \, \un B}{}^{k l}$).  With the forms of (\ref{eq:35}) and
(\ref{eq:36}) together with the known spectrum of 4D, $N$ = 4
supergravity, we need only find consistent solutions of the superspace
Bianchi identities.    The quantity ${\cal T}_{k l} $ denote the
generators of a rank six gauge group.

The quantity $ {\cal E}_{i j}{}^{k l}$ that appears in the
spinor-spinor component of the super 2-form field strength ($F_{\a i
\, \b j}{}^{k l} = C_{\a \b} \, {\cal E}_{i j}{}^{k l}$) was
required by the Bianchi identities to be an invertible 6 $\times$
6 matrix and given the spectrum of component fields had to be of
the form \cite{R15}
\be  \eqalign{
{\cal E}_{i j}{}^{k l} ~=~ \d_i {}^{[k }\, \d_j {}^{l] } \,
U(W) ~+~ \e_{i j} {}^{k l } \, V(W)
~~~.
} \label{eq:37} \ee
This is the most general form that is consistent with SO(4) symmetry
and it is here that this assumption enters the analysis in a forceful
manner.  Once (\ref{eq:36}) and the spectrum of 4D, $N$ = 4 supergravity
are  used as inputs for solving the superspace Bianchi identities to
which (\ref{eq:36})  is subject, there emerges the condition
\be  \eqalign{
| U |^2 ~-~ |V|^2 = \left(\begin{array}{ccc}
  ~~1\\
  ~~0\\
  ~~-\, 1\\
\end{array}\right)
~~~,
} \label{eq:38} \ee
that we call the ``modulus constraint'' and whose uniqueness was
verified in the first and second work of \cite{R16}.  In (\ref{eq:37})
$W$ is a chiral superfield whose leading components are $\cal Z$ from
(\ref{eq:23}).  Under these results all superspace Bianchi identities
were found to be satisfied.

 From the view of supergeometry, (almost) the only freedom that one has
for 4D, $N$ = 4 supergravity is in the ``modulus choice'' and the
rank six gauge group choice.  Some  choices (but not exhaustively)
are\footnote{The choice ${\cal Z}^{(3)} \otimes$ $SU(2)$ may be
regarded as a degenerate case of the $SU(2) \otimes$ $SU(2)$
\newline ${~~~~~}$ theory wherein one of the non-Abelian coupling
constants is set to zero, e.\ g.\ \cite{R19}.}; U${}^{(6)}$(1),
$SO(4)$ and $SU(2) \otimes$ $SU(2)$.  These observations provide a
``matrix classification scheme'' for understanding the
versions of 4D, $N$ = 4 supergravity.  Along a vertical axis,
we list the choice of gauge group. Along a horizontal axis, we list
the modulus choice.  This leads to a simple taxonomy.

\vspace{0.1cm}
\begin{center}
\centerline{{\bf {Taxonomy of 4D,}} ${\bf {N = 4}}$ {\bf {
Supergravity}} }
\vskip.1cm
\begin{tabular}{|c|c|c|c|}\hline
${\rm {~~~~~~}} $ & $~~+1 ~~$ & $ ~~0~~
$  & $~~- 1~~ $ \\
\hline
$ ~{\cal Z}^{(6)}~$ & \cite{R9,R10,R13} & \cite{R11} & \cite{R18,R17}
\\
\hline
$ ~SO(4)~$ & \cite{R13}& $ ~--~$  & $ ~ * ~$ \\ \hline
$ ~SU(2) \otimes SU(2)~$ & \cite{R18} & \cite{R14}  & $ ~
* ~$ \\
\hline
\end{tabular}
\end{center}
\vspace{0.1cm}
\centerline{{Table I}}
The numerical entries in this table indicate in which reference the
construction was carried out, $--$ indicates the impossibility to
carry out such a construction and $*$ indicates the model has simply
not been explicitly presented.  The modulus choice = +1 and gauge
group = $SU(2) \otimes SU(2)$ model in \cite{R18} was the first
extended supergravity model presented in the literature that allows
de Sitter space as a spontaneously broken phase.

The modulus choice is important for a number of reasons.  Perhaps the
most important of these is that this choice will ultimately control the
number of spin-0 fields that can appear in the potential after gauging
the automorphism group or sub-groups thereof.  For each of the moduli
choices, one can find an explicit functional relation expressing $U$
and $V$ in terms of $W$.  These take the forms;
\be  \eqalign{
&U ~=~ {1 \over {\sqrt {1 ~-~ | W|^2 ~}}} ~~,~~~~~~~~~~ V ~=~ {W \over
{\sqrt {1 ~-~ | W|^2 ~ }}}  ~~~,  \cr
&U ~=~ {1 \over {\sqrt {1 ~-~ W ~-~ {\Bar W}~ }}} ~~~,~~~ V
~=~ {1 \over {\sqrt {1 ~-~ W ~-~ {\Bar W}~ }}} ~~~,  \cr
&U ~=~  {W \over {\sqrt {1 ~-~ | W|^2 ~}}}  ~~~,~~~
~~~~~~ V  ~=~ {1 \over {\sqrt {1 ~-~ | W|^2 ~}}}   ~~~, \cr
} \label{eq:39} \ee
respectively for the three cases indicated in (\ref{eq:38}).  Note that
for the modulus choice of $\pm 1$, two scalar functions (i\ e.\ ${\cal
R}e(W)$ and ${\cal I}m(W)$) appear in ${\cal E}_{i j}{}^{k l}$.  For
the modulus choice of 0, only one scalar field appears (i\ e.\ ${\cal
R}e(W)$) in ${\cal E}_{i j}{}^{k l}$.

In a paragraph above, the word ``almost'' appeared in a parenthetical
remark.  The reason is because the Nicolai-Townsend model \cite{R12}
is not among the classification above.  It was later shown in the
third work of \cite{R16} that this version of 4D, $N$ = 4 supergravity
exist in a superspace formulation where the complex chiral superfield
$W$ is replaced by a real scalar superfield ${\cal V} \equiv  W +
{\Bar W}$ and this is only consistent for the modulus choice of zero.
Thus, the superspace geometry of the Nicolai-Townsend formulation is
described by the real superfield ${\cal V}$ and it is convenient to
relate this superfield to the component ``dilaton'' in the theory
$\varphi(x)$  via ${\cal V} = 1 - exp(-\varphi)$.  In addition to the
superspace  supercovariant derivative (\ref{eq:35}), it is necessary
to introduce a super 2-form gauge superfield $B_{\un A \un B}$ and
its associated super 3-form field strength $H_{\un A \un B \un C}$
for a complete supergeometrical description.  This exhaust all freedom
in superspace to describe the 4D, $N$ = 4 supergravity models.

A final point about the modulus choice is that it is literally a
choice of moduli (in the technical sense of the word) that enter the
theory.  Let us see how this works in some more detail.  Due to the
gauge invariance of the two-form in the Nicolai-Townsend formulation
and via the duality transformation described by (\ref{eq:32}), we can
be assured that the field $B(x)$ only enters ${\cal S}_{ {\rm {SU(2)}}
\otimes {\rm {SU(2)}}} $ through its derivative.  In turn this means
that its action must admit the symmetry generated by $B \to B
+ c_0$ for an arbitrary constant $c_0$.  Thus we have derived the fact
that the theories which appear in the middle column of our taxonomy
possess a space of moduli for their $B$-fields and such moduli
do {\it {not}} appear for the other member of the classification.

To gauge the $SU(2) \otimes SU(2)$ group requires two coupling constants
$g_1$ and $g_2$.  The second work of \cite{R16} established that the two
coupling constants are not independent but are related by a
{\bf Z}${}_2$-valued parameter
\be  \eqalign{
g_1 ~=~ g  ~~~,~~~ g_2 ~=~ \eta \, g ~~,~~ \eta ~=~ \pm 1 ~~~,
} \label{eq:40} \ee
and when combined with the results of \cite{R18}, where it was shown
that the model develops a spin-0 potential function of the form
(with $\eta_{\pm} \equiv 1 + \eta$), leads to
\be  \eqalign{ {~~~~~}
P(W, \, {\Bar W}) &=~  \frac 12 \, g^2  \,(~ 3 \, |\, \eta_+ \,
U ~+~ \eta_- \, {\Bar V} \, |^2 ~-~ | \, \eta_+ \, V ~+~ \eta_-
\, {\Bar U} \, |^2  ~)  \cr
&=~   \, g^2  {\Big [ } ~  { { ~\eta_+ (\, 3 ~-~  |W |^2 \, ) ~-~
\eta_- (\, 1 ~-~ 3 |W |^2  \,) ~} \over  {1 ~-~  | W|^2} }  ~ \Big]
~~~. }
\label{eq:41} \ee
When $\eta_- = 0$, this potential at its critical point describes a
cosmological constant $\l_{AdS} = - 6 g^2$ appropriate for an anti-de
Sitter geometry.  When $\eta_+ = 0$, this potential at its critical
point describes a cosmological constant $\l_{dS} = 2 g^2$ appropriate for
an de Sitter geometry (clearly $\l_{AdS}/\l_{dS} = - 3$).

The discovery of a mechanism that allowed for de Sitter backgrounds
in the context of 4D, $N$ = 4 extended supergravity models triggered
an extensive program of study for generalizations to higher values of
$N$.  This effort, led by Hull, Hull and Warner \cite{R20} and
others, continues even to this day \cite{R21}.  The importance of these
works can be judged against the prior construction of gauged SO(8)
4D, $N$ = 8 models \cite{R22} in which de Sitter space was ruled out.

\section {Supersymmetry Breaking and a Nonvanishing
Cosmological Constant}

~~~~In the same period of time that we were investigating the appearance
of de Sitter backgrounds in extend supergravity, we also presented a
discussion \cite{R23} about a linking between the appearance of a
spacetime background of constant curvature and the breakdown of rigid
supersymmetry in the context of 4D, $N$ = 1 supergravity theories.
This linkage was first noted in {\it {Superspace}} \cite{R24}, where
it was observed that the choice of auxiliary fields required of an
off-shell supergravity multiplet is sensitive to the presence of
background spaces of constant curvature.  For some sets of auxiliary
fields, the mere presence of any non-vanishing cosmological constant
implies the breaking of global supersymmetry!   Since this result
seems largely unknown to the  community, we will expend an effort in
the following to explain why this result is obtained.

As is well known, 4D, $N$ = 1 supergravity describes the propagation of
spin-2 and spin-3/2 degrees of freedom, e${}_{\un a}{}^{\un m}(x)$
and $\psi{}_{\un a}{}^{\a}(x)$.  However, the consistency of the local
supersymmetry algebra implies that if {\it {only}} these fields are
introduced, then they must satisfy some equations of motion.  In order
to avoid this restriction, auxiliary fields are required.  The set of
auxiliary fields needed for this is {\it {not}} unique.  One such set
consists of ($B, \, {\Bar B},\, A_{\un a}$) and these are called the
``minimal set of auxiliary fields'' as discovered by van Nieuwenhuizen
and Ferrara \cite{R25}.  A different set, called ``non-minimal set of
auxiliary fields,''  consists of ($\l_{\a}, \, B,
\, {\Bar B},\, A_{\un a}, \, w_{\un a},\, {\Bar w}{}_{\un a}, \,
\chi_{\a},$) and had previously been implied by the work of
Breitenlohner \cite{R26}.  Each of these sets of auxiliary fields
correspond to distinct supergeometries.  For the minimal set, this
geometry is expressed in term of the superfields $W_{\a \b \g}$,
$G_{\un a}$ and $R$ are known as the ``irreducible supergravity field
strengths.''  While in the non-minimal set the geometry is
expressible in terms of the superfields $W_{\a \b \g}$, $G_{\un a}$ and
$T_{\a}$ are  the ``irreducible supergravity field strengths.''

The superspace commutator algebra of the minimal supergravity
covariant derivative is given by
$$  \eqalign{ {~~~~}
[ \nabla_{\a }, \nabla_{\b} \}  &=~ - 2 \Bar R \,{ \cal M}_{\a \b}
~~~~~, ~~~~~
[ \nabla_{\a }, { \Bar {\nabla}}_{\dot \a} \}  ~=~ i {\nabla}_{ \un a}
~~~~~, \cr
[ \nabla_{\a }, \nabla_{\un b} \}  &=~ - i \, C_{ \a \b} [~ \Bar R \,
{\Bar {\nabla}}_{\dot \b} ~-~ G^{\g} {}_{\dot \b } \nabla_{\g} ~ ] ~-~
i \, ( {\Bar \nabla}_{\dot \b} \Bar R )  { \cal M}_{\a \b} ~~ \cr
&~~~~~+~ i \, C_{\a \b} [~ {\Bar W}_{\dot \b \dot \g} {}^{\dot \d}
  { \Bar {\cal M}}_{\dot \d} {}^{\dot \g} ~-~ ( \nabla^{\d} G_{\g \dot \b})
{\cal M}_{\d}  {}^{ \g}  ~ ]  ~~~, \cr
[ {\nabla}_{\un a }, \nabla_{\un b} \}  &=~ \{ ~ i \fracm12
C_{\a \b} G^{\g} {}_{ ( \dot \a} \nabla_{\g \dot \b)}    \cr
&~~~~~~+~ [~ C_{\dot \a \dot \b} W_{\a \b}{}^{ \g} ~+~ \fracm 12 C_{\a \b}
({\Bar \nabla}_{( \dot \a} G^{\g}{}_{ \dot \b )}) ~-~ \fracm 12 C_{\dot \a
\dot \b} \,( \nabla_{ ( \a} R  ) \,\d_{\b)}{}^{\g} ~ ] \nabla_{\g} } $$
\be  \eqalign{ {~~~~}
&~~~~~~-~ [~ C_{\dot \a \dot \b} W_{ \a \b \g \d} ~-~ \fracm 12 C_{\a
\b} \, ({\Bar \nabla}_{( \dot \a} \nabla_{\g} G_{\d}{}_{ \dot \b )})~]
{\cal M}^{\g \d}   \cr
&~~~~~~- i \fracm 18 C_{\dot \a \dot \b} C_{\g ( \a}\,  [\,
\nabla_{\b ) } {}^{\dot \e} G_{\d \dot \e} ~+~  \nabla_{\d}{}^{\dot \e}
G_{\b ) \dot \e} \, ] \,  {\cal M}^{\g \d} \cr
&~~~~~~- \fracm 12 C_{ \dot \a \dot \b } C_{\g ( \a } C_{\b ) \d } \,
[ \, {\Bar \nabla}^2 \Bar R ~+~ 2 R \Bar R ~\, ] {\cal M}^{\g \d}
~ \} ~+~  {\rm {h.\, c.}}  ~~~,
} \label{eq:42} \ee
The superfield $R$ is subject to the superdifferential equations
\be \eqalign{
{\Bar {\nabla}}_{\dot \b} R ~=~  0 ~~~,~~~ \nabla^{\a} R ~+~
{\Bar {\nabla}}_{\dot \a}  G^{\a \dot \a} ~=~ 0  ~~~.
} \label{eq:43} \ee
We now wish to consider several limits of the equations in (\ref{eq:42})
and (\ref{eq:43});

Limit A. We set $W_{\a \b \g} = 0$, $G_{\un a} = 0$ and $R = 0$
and see that this limit is \newline ${~~~~~~~~~~~~~~~~}$ consistent
with the differential equations in (\ref{eq:43}) and that (\ref{eq:42})
\newline ${~~~~~~~~~~~~~~~~}$ agrees precisely with (\ref{eq:15}) if
we set $\ell_{\cal F} = 0$ in the latter.  This is
\newline ${~~~~~~~~~~~~~~~~}$ the flat Minkowski limit
of the curved supergravity superspace.

Limit B. We set $W_{\a \b \g} \neq 0$, $G_{\un a} = 0$ and $R = 0$
and see that this limit is \newline ${~~~~~~~~~~~~~~~~}$ consistent
with the differential equations in (\ref{eq:43}) and that (\ref{eq:42})
\newline ${~~~~~~~~~~~~~~~~}$ implies a non-trivial spacetime curvature.
This superspace de- \newline ${~~~~~~~~~~~~~~~~}$ scribes the
``on-shell'' limit with vanishing auxiliary fields in \cite{A1}.

Limit C. We set $W_{\a \b \g} \neq 0$, $G_{\un a} = 0$ and $R = \ell_{\cal
F}$  and see that this limit is \newline ${~~~~~~~~~~~~~~~~}$ consistent
with the differential equations in (\ref{eq:43}) and that (\ref{eq:42})
\newline ${~~~~~~~~~~~~~~~~}$ implies a non-trivial spacetime curvature.
This  superspace de- \newline ${~~~~~~~~~~~~~~~~}$ scribes the
``on-shell'' supergravity theory of \cite{R8} in an anti-de
\newline ${~~~~~~~~~~~~~~~~}$ Sitter background.

Limit D. We set $W_{\a \b \g} = 0$, $G_{\un a} = 0$ and $R = \ell_{\cal
F}$ and see that this limit is \newline ${~~~~~~~~~~~~~~~~}$ consistent
with the differential equations in (\ref{eq:43}) and that (\ref{eq:42})
\newline ${~~~~~~~~~~~~~~~~}$ agrees precisely with (\ref{eq:15}).
This is the flat anti-de Sitter limit of
\newline ${~~~~~~~~~~~~~~~~}$ the curved supergravity
superspace.

We now wish to repeat this analysis and show that the supergeometry of
the Breitenlohner auxiliary fields behaves in a drastically different
manner in the some of limits above.  The non-minimal superspace geometry
can be cast in the form
\be \eqalign{
[ \, \nabla_{\a} ~,~ \nabla_{\b} \,  \} &=~ \fracm 12 \, T_{(\a}
\, \nabla_{\b)} ~-~ 2 \, ( {\Bar R} \,+\, T^{\a} T_{\a} )
  {\cal M}_{\a \b} ~~~,  \cr
[ \, \nabla_{\a} ~,~ {\Bar \nabla}{}_{\Dot \b} \,  \} &=~ i \,
\nabla_{\un a} ~-~   \fracm 12 \, ( \, T_{\a} \, {\Bar \nabla}
{}_{\Dot \b}  \, + \,  {\Bar T}{}_{\Dot \a} \,
\nabla_{\b} \,)  \cr
[ \, \nabla_{\a} ~,~ {\nabla}{}_{\un b} \,  \} &=~ i \, C_{\a \b}
\, ( {\Bar R} \,+\, T^{\a} T_{\a} ){\Bar \nabla}{}_{\Dot \b}
~+~ i C_{\a \b} G^{\a}{}_{\b} \nabla_{\g}  \cr
&~~~~-~ i \fracm 12 \, ( \, \nabla_{\b} \, {\Bar T}{}_{\Dot \b} \,
  - \, {\Bar \nabla}{}_{\Dot \b}  \,T_{\b} \, +\, T_{\b} \, {\Bar
T}{}_{\Dot \b} \, ) \, \nabla_{\a}  \cr
&~~~~-~ i \, [ \, ( \,  {\Bar \nabla}{}_{\Dot \b} \, - \,
{\Bar T}{}_{\Dot \b} \, ) {\Bar R} \, ]  \,
{\cal M}_{\a \b} ~-~ i C_{\a \b} ( \nabla^{\g} G_{\d \Dot \b} )
\, {\cal M}_{\g}{}^{ \d} \cr
&~~~~+~ i C_{\a \b} \, [ \,\, {\Bar W}{}_{\Dot \b \Dot \d}{}^{
\Dot \g} \, {\Bar {\cal M }}{}_{\Dot \g}{}^{ \Dot \d} \, - \,
\fracm 13 (\, \nabla^{\g} G_{\g \Dot \d}  \cr
&~~~~~~~~~~~~~~~~~-~ ( \,  {\Bar \nabla}{}_{\Dot \a} \, - \,
{\Bar T}{}_{\Dot \a} \, ) \, ( {\Bar R} \,+\, T^{\a} T_{\a})
~)  {\Bar {\cal M }}{}_{\Dot \b}{}^{ \Dot \d} ~] ~~~,
}\label{eq:44} \ee
where for simplicity we have omitted the explicit form of
$ [  \nabla_{\un a} \, , \, {\nabla}{}_{\un b}  \} $.  But this
may be found from
\be \eqalign{
  [ \, \nabla_{\un a} ~ , ~ {\nabla}{}_{\un b} \, \} &=~ [ ~ - i
\, ( [ \nabla_{\a} \, , \, [  {\Bar \nabla}{}_{\Dot \a} \, , \,
{\nabla }{}_{\un b} \} \,\} ~-~ T_{\a} \,  [  \nabla_{\a} \, , \,
{\nabla}{}_{\un b}  \}  \cr
&~~~~~~~~~~~~~~-~ ( \, {\nabla}_{\un b}\, {\Bar T}{}_{\Dot \a})
\, \nabla_{\a} ~+~ {\rm {h. \, c.}} ~] ~~~,
}\label{eq:45} \ee
In particular in this theory $R$ is subject to the superdifferential
equations
\be \eqalign{  {~~}
{\Bar {\nabla}}_{\dot \b} R ~=~  0 ~~~,~~~
(\nabla^{\a} ~-~ \fracm 12 T_{\a} ) R ~+~ {\Bar {\nabla}}_{\dot \a}
G^{\a \dot \a}  ~=~ 0  ~~~,~~~
{\Bar R} ~=~ - \fracm 12 \, \nabla^{\a} T_{\a}
   ~~~,}
  \label{eq:46}  \ee
Repeating the same limits as before but now investigating the equations in
(\ref{eq:44}), (\ref{eq:45}) and (\ref{eq:46});

Limit A. We set $W_{\a \b \g} = 0$, $G_{\un a} = 0$ and $T_{\a} = 0
\to R = 0 $
and see that this \newline ${~~~~~~~~~~~~~~~~}$ limit is consistent
with the differential equations in (\ref{eq:46}) and that
\newline ${~~~~~~~~~~~~~~~~}$ (\ref{eq:44}) and (\ref{eq:45}) agree
with (\ref{eq:15}) if  we set $\ell_{\cal F} = 0$ in the latter.
This \newline ${~~~~~~~~~~~~~~~~}$ is the flat Minkowski limit
of the curved supergravity superspace.

Limit B. We set $W_{\a \b \g} \neq 0$, $G_{\un a} = 0$ and $T_{\a} = 0
\to R = 0$ and see that this \newline ${~~~~~~~~~~~~~~~~}$ limit is
consistent  with the differential equations in (\ref{eq:46}) and that
\newline ${~~~~~~~~~~~~~~~~}$ (\ref{eq:45}) implies a non-trivial
spacetime curvature.  This superspace de- \newline ${~~~~~~~~~~~~~~~~}$
scribes the ``on-shell'' limit with vanishing auxiliary fields in
\cite{A1}.

Limit C. We set $W_{\a \b \g} \neq 0$, $G_{\un a} = 0$,  $T_{\a} = 0,
\, R = \ell_{\cal F}$ and see that this \newline
${~~~~~~~~~~~~~~~~}$ limit is {\it {not}} consistent with the differential
equations in (\ref{eq:46}).

Limit D. We set $W_{\a \b \g} = 0$, $G_{\un a} = 0$, $T_{\a} = 0,
\, R = \ell_{\cal F}$ and see that this  \newline
${~~~~~~~~~~~~~~~~}$ limit is {\it {not}} consistent with the differential
equations in (\ref{eq:46}).

\noindent
In particular, it is the final equation in (\ref{eq:46}) that is always
inconsistent.

This phenomenon was also investigated in term of the compensating field
formalism \cite{R24} and that study also supported this assertion.  In
particular, there is an important technical difference between the manner
in which the chiral compensator superfield $\varphi$ (see \cite{R28})
and the linear compensator $\U$ (see also \cite{R28}) enter their
respective supergravity actions.  The chiral compensator enters the
supergravity action in such a way that the sign of its kinetic term
is {\it {opposite}} to that of a matter chiral superfield.  On the other
hand, the sign of the linear compensator enters the ($n$ = - 1) supergravity
action in such a way that the sign of its kinetic term  is the {\it {same}} as
that of a matter linear superfield.  Thus, we  long ago reached the conclusion
that {\it {all}} spaces of constant curvature break rigid supersymmetry, if the
off-shell non-minimal formulation of supergravity is required.

This behavior may not be an academic matter, especially if the
``P-term inflation model'' recently suggested by Kallosh \cite{R27}
is thought to provide a way in which to reconcile quintessence and
supersymmetry in a ``stringy''  context.   {\it {All}} known $N$ = 2
off-shell supergravity multiplets possess a  subsector $N$ = 1 off-shell
supergravity multiplet whose auxiliary  field {\it {are}} the
Breitenlohner set.

One final point to note about this phenomenon is that it may well
provide an example of a model in which the breaking of flavor symmetry
results in the breaking of supersymmetry.  The point is that usually
in breaking internal flavor symmetries most such models generate a
cosmological term.  But such a term for this supergravity theory must
necessarily drive supersymmetry breaking so that the two would be
intimately linked. Thus in these models, electroweak breaking,
supersymmetry breaking {\it {and}} the cosmological constant are
all related.

\section {Unexplored Issues for de Sitter Space in $N$ = 8 Supergravity}

~~~~The works of de Wit and Nicolai \cite{R22} and as well Hull et.\ al.\
\cite{R20} (as well as subsequent works based upon this) begin at a
starting point of 4D, $N$ = 8 supergravity where all of the 70 spin-0
fields are represented by scalars.  If there were no other options, then
one might not raise the issue of additional presently unknown gauged 4D,
$N$ = 8 supergravity models. There are other options.

Even in the original construction of 4D, $N$ = 8 supergravity, Cremmer and
Julia \cite{R29} were very clear about this issue.  At a certain point in
their derivation of the 4D, $N$ = 8 supergravity action from 11D, $N$ = 1
supergravity action, it is required to  perform a duality transformation
from a set of seven 2-forms to a set of seven scalars\footnote{These same
seven 2-forms will appear below in our discussion starting from the 10D,
$N$ = 2A \newline ${~~~~~}$ supergravity theory.}.

In the conventional 4D, $N$ = 8 superspace supergravity theory, there is
also the analog of ${\cal E}_{i j}{}^{k l}$ taking the form,
\be \eqalign{
{\cal E}_{i j}{}^{k l} ~=~  U_{i j} {}^{k l} (\Phi)\,
~+~ V_{i j} {}^{k l }(\Phi)  ~~~,
}  \label{eq:47}  \ee
which depends on the irreducible 4D, $N$ = 8 superspace supergravity field
strength
$\Phi_{i j k l}$.  This latter quantity is the  4D, $N$ = 8 analog
of $W$ and must be totally antisymmetric satisfying,
\be \eqalign{
(\Phi_{i j k l})^* ~=~ \pm \fracm {1\,}{4!} \, \e^{i j k l r s t u}
\, \Phi_{r s t u} ~~~.
}  \label{eq:48}  \ee
Finally, the functions $U_{i j} {}^{k l}$ and $V_{i j} {}^{k l}$ were
chosen according to the following prescription.

Since the quantity $\Phi_{i j k l}$ is complex, so too must be $U_{i j}
{}^{k l}$ and $V_{i j} {}^{k l}$. They therefore possess
complex conjugates ${\Bar U}{}^{i j} {}_{k l}$ and ${\Bar V}{}^{i j}
{}_{k l}$.  These four functions can be assembled into a matrix ${\cal B}$
defined by
\be \eqalign{
{\cal B} ~\equiv~ \left(\begin{array}{cc}
~U_{i j} {}^{k l} & ~~ V_{i j} {}^{k l }  \\
{}~&~\\
~ {\Bar V}{}^{i j} {}_{k l} & ~~  {\Bar U}{}^{i j} {}_{k l} \\
\end{array}\right) ~~~,
}  \label{eq:49}  \ee
and the functional dependence of $U_{i j} {}^{k l}$ and $V_{i j} {}^{k l}$
upon $\Phi_{i j k l }$ is fixed by the condition that
\be \eqalign{
{\cal B} ~=~ exp\left(\begin{array}{cc}
~0 & ~~ \Phi_{i j k l }  \\
{}~&~\\
~ {\Bar \Phi}{}^{i j k l }  & ~~ 0 \\
\end{array}\right) ~~~.
}  \label{eq:50}  \ee
The 70 scalar fields that enter the functions $U_{i j} {}^{k l}$ and
$V_{i j} {}^{k l}$ possess no moduli.  That is, the function ${\cal
E}_{i j}{}^{k l}$ is {\it {not}} invariant under $\Phi_{i j k l }
\to \Phi_{i j k l } + c_{i j k l }$ for any choice of constants
$c_{i j k l }$.  The theory has a zero dimensional moduli space.

The toroidal compactification of type-II supergravity theories provides
alternatives.  Since both the type-IIA and type-IIB theories exist, there
are guaranteed to exist at least two alternative 4D, $N$ = 8 theories that
we refer to as the 4D, $N$ = 8A and 4D, $N$ = 8B theories. In particular,
the group Spin(6) seems to play the role  of organizing the 4D, $N$ = 8
supergravity fields into its representations. Let us present the toroidal
reduction in the form of a table.

\begin{center}
\centerline{{{\bf {D = 10,~ $N$ = 2A}}} {{\bf Supergravity
Reduction}}}
\renewcommand\arraystretch{1.2} \vspace{0.1cm}
\begin{tabular}{|c|c|c|c| }\hline
$~~$ D = 10  $~~$ & $~~~$ D = 4 $~~~$ & $~$ Multiplicity $~$
& $~$ 4D Spin $~$  \\ \hline\hline
  $  {\hat e}_{\hat \m} {}^{\hat m}     $ & $  e_{\un m} {}^{\un a},
~ \varphi_{(\hat \a \hat \b)},~ \varphi, ~ {\tilde A}_{\un m}{}^{\hat \a},
~~~ ( \varphi_{\hat \a}{}^{ \hat \a} = 2\varphi)$  & $1, ~
20, ~ 1, ~ 6$ & 2, 0, 0, 1\\ \hline
  $ {\hat \psi}_{\hat m} $ & $ \psi_{\un m},~  \psi^{\hat \a},~
    \psi ,~~~~ (\G^{\hat \a} \psi_{\hat \a} = 0)  $ & $8, ~
40, ~ 8$ & 3/2, 1/2, 1/2\\  \hline
  $  {\hat \chi}        $ & $  {\chi}  $ & $8$ & 0 \\ \hline
$ {\hat A}_{\tilde \m}  $ & $ {\tilde A}_{\un m},~{\varphi}_{\hat \a}$
  &  $1, ~ 6$    & 1, 0  \\ \hline
$ {\hat B}_{\hat \m \hat \n}  $ & $ {B}_{\un m \un n},~
  {A}_{ \un m \hat \a},~{\varphi}_{[\hat \a \hat \b]}$
  & $1, ~ 6, ~ 15$  & 0, 1, 0 \\  \hline
$ {\hat A}_{\hat \m \hat \n \hat \r}  $ & $ {B}_{\un m \un n \hat \a},
~{A}_{\un m \hat \a \hat \b},~  {\varphi}_{\hat \a \hat \b \hat \g}$
& $6,~15, ~20$ & 0  1, 0 \\  \hline
$ {\hat \phi}  $ & $ \phi $ &  $1$ & 0
  \\  \hline
\end{tabular}
\end{center}
\vspace{0.1cm}
\centerline{{Table II}}

In particular, there are some points of note.

  (a.) The eight gravitini (and dilatini) in the theory may
       \newline ${~~~~~~~~~}$ be regarded as forming the
       spinor representation of \newline ${~~~~~~~~~}$
       Spin(6).  All multiplicities fall into Spin(6) representa-
       \newline ${~~~~~~~~~}$ tions.

  (b.) The spin-one fields (whose superspace ${\cal E}_{i j}
       {}^{k l}$-functions
       \newline ${~~~~~~~~~}$ determine the how the spin-0 fields
       appear) are \newline ${~~~~~~~~~}$
       in multiplicities of 1 + 6 + 6 + 15 which is exactly
       \newline ${~~~~~~~~~}$ what is needed for the gauge
       group

${~~~~~~~~~}$  $SO(2) \otimes SO(4) \otimes SO(4) \otimes SO(6)$.

  (c.) There are seven 2-forms, ${B}_{\un m \un n}$ and
       ${B}_{\un m \un n \hat \a}$, which may after
       \newline ${~~~~~~~~~}$ dualization
       yield a maximum of seven spin-0 fields that
       \newline ${~~~~~~~~~}$ possess a seven dimensional
       moduli space.

A rather similar analysis can be performed on the type-IIB theory.
We note that the fermion results here are unchanged so in order to
simplify our considerations we neglect showing the fermions.  We
once again resort to a table.

\begin{center}
\centerline{{$\bf D = 10,~ N = 2B$ {\bf {Supergravity Reduction}}}}
\renewcommand\arraystretch{1.2}  \vspace{0.1cm}
\begin{tabular}{|c|c|c|c| }\hline
$~~~$ D = 10  $~~~$ & $~~~$ D = 4 $~~~$ & $~$ Multiplicity $~$
  & $~~~$ 4D Spin $~~~$ \\ \hline\hline
$  { e}_{\un a} {}^{\un m}     $ & $
\left(\begin{array}{cc}
~{\hat e}_{\un a} {}^{\un m} & ~~ A_{\un a} {}^{\hat m} \\
{}~&~\\
~0 & ~~\D_{\hat a} {}^{\hat m} \\
\end{array}\right) $   & $
\left(\begin{array}{cc}
1 & ~~ 6 \\ {}~&~\\ ~0 & ~~21 \\
\end{array} \right) $    & $
\left(\begin{array}{cc}
2 & ~~ 1 \\ {}~&~\\ ~0 & ~~0 \\
\end{array}\right) $   \\ \hline
$ G(B)_{\un a \un b \un c}  $ & $ G(B)_{\un a \un b \un c},~ $
$G(B)_{\un a \un b \hat c}     ,~G(B)_{\un a \hat b \hat c}$
& $1,~ 6, ~ 15$   & 0, 1, 0 \\  \hline
$ {\Phi}  $ & $ \Phi $ & $1$  & 0 \\  \hline
$ { A}  $ & $  A $ & 1  & 0  \\  \hline
$ F(A)_{\un a \un b \un c}  $ & $ F(A)_{\un a \un b \un c},~
F(A)_{\un a \un b \hat c}  ,~F(A)_{\un a \hat b \hat c}$
  & $1,~ 6, ~ 15$   & 0, 1, 0 \\  \hline
$ F(A)_{\un a \un b \un c \un d \un e}  $ & $ F(A)_{\un a \un b
\hat c \hat d \hat e },~F(A)_{\un a \hat b \hat c \hat d \hat e}$
  &  $10,~15$    & 1, 0 \\  \hline
\end{tabular}
\end{center}
\vspace{0.1cm}
\centerline{{Table III}}

Similar to the last case, there are three points of note.

  (a.) The multiplicities fall into Spin(6) representations.

  (b.) The spin-one fields are in multiplicities of 6 +
       6 + 6  + 10.

  (c.) There are two 2-forms field strengths, $G_{\un a
       \un b \un c}$ and $F_{\un a \un b \un c}$, \newline
       ${~~~~~~~~~}$ which, after dualization, yield two
       spin-0 fields that \newline ${~~~~~~~~~}$ could
       possess a two dimensional moduli space.

Since there is no action for the IIB theory in 10D, there is a
subtlety in counting the total number of moduli.  In addition to
the moduli that result from the 2-forms, it might be argued that
additional moduli could emerge for some of the other scalar fields
that are contained in the table above.  To understand this possibility
let us consider the multiplicities of all 70 spin-0 fields.

If we assume that some 4D action exists that respects the Spin(6)
symmetry of the fields, then the only dimensions of moduli space
less than seven that could arise are two, three or four.   Moreover
there is {\it {no}} choice, respecting the  Spin(6) representation of the
scalar fields, by assuming that moduli can appear arbitrarily for the
scalar fields, that leads to either six or seven.  This argument suggest
that the number of moduli can {\it {never}} be equal to either
six or seven.

We thus argue based on the dimensions of the different moduli spaces,
respecting the symmetry of the spin-0 fields, that all this evidence points
toward the existence of a minimum of {\it {three}} distinct 4D, $N$ = 8
supergravity theories!  We follow the lead of our $N$ = 4 taxonomy listing
the moduli choices horizontally and the gauge group choices vertically.

\vspace{0.1cm}
\begin{center}
\centerline{
{\bf {Putative and Partial Taxonomy of 4D,}} $\bf N$ {\bf {= 8
Supergravity}} }
\vspace{0.1cm}
\begin{tabular}{|c|c|c|c|}\hline
${\rm {~~~~~~}} $ & $~~N = 8 ~~$ & $ ~~N = 8A~~
$  & $~~N = 8B~~ $ \\
\hline
$ ~{\cal Z}^{(28)}~$ & \cite{R29} & $ ~ * ~$ & $ ~ * ~$ \\
\hline
$ ~SO(8)~$ & \cite{R22} & $ ~ * ~$  & $ ~* ~$ \\
\hline
\end{tabular}
\end{center}
\vspace{0.1cm}
\centerline{{Table IV}}

We know that this table is partial precisely due to the Hull-type
constructions.  In addition to the gauge groups listed above, all
those explored in the works of \cite{R20} and their descendants may
be added to the first column.  Colloquially, we may say that studies
following on \cite{R20} simply move us down the first column.  There
has never been a complete construction of the other putative theories
suggested above.   Similarly, there has never been a complete
supergeometrical analysis for the $N$ = 8 case as has been done for
the $N$ = 4 theories in \cite{R16}.  The question of whether a
successfully gauged version of either the $N$ = 8A theory or $N$
= 8B theory and which leads to models that are distinct from those
already elucidated by Hull's approach seems to be a worthwhile one
to investigate.

The 4D, $N$ = 8 story may not end with just the observations that
we have made so far.  A number of years ago \cite{R30}, we noted
that a 4D, $N$ = 8 Green-Schwarz construction suggests an even
larger zoo of theories.   The GS action is well known,
$$ \eqalign{
S_{GS}&=~\int d^2 \s V^{-1}\Bigl [ ~ -  \P_{\pp}~^{\un a}
\P_{\mm~\un a}~+~ \int_{0}^{1}dy \hat {\P}_{y}~^{\un C}\hat{\P}_{
\pp}~^{\un B}\hat{ \P}_{\mm}~^{\un A}\hat{G}_{\un A\un B\un C} ~
\Bigr ]~~~, ~~~~}$$
$$\P _{\pp}~^{\un A}~=~V_{\pp}~^{ m}( \pa_{m}Z^{\un M}) E_{\un M
}~^{\un A}~~,~~ \P _{\mm}~^{\un A}~=~V_{\mm}~^{m}(\pa_{m}Z^{\un
M}) E_{\un M}~^{ \un A} ~~~,$$
\be \eqalign{ {~~~~~~}
\hat{Z}^{\un M}~=~Z^{\un M}(\s,\t,y)~~,~~\hat{\P}_{y}{}^{\un
A}~=~(\pa_{y}\hat{Z}^{\un M})E_{\un M} {}^{\un A} ({\hat Z})~~,~~
\hat{G}_{\un A\un B\un C} = G_{\un A\un B\un C}(\hat{Z}) ~~~,
}  \label{eq:51}  \ee
where we refer the reader to \cite{R30} for notational conventions.
However, instead of interpreting this expression in some higher
dimension, we proposed to study some of its properties in 4D.  In
particular we let $Z^{\un M} \equiv  (\Theta^{\m ~ i}, \Theta^{\m ~
i'}, {\bar {\Theta}}^{\dot \m}{}_{i}, {\bar {\Theta}}^{\dot \m}{
}_{i'} , X^{\un m })$ and define the supercoordinate of the
string where $ {X}^{\un m}(\t, \s)$ is a four dimensional bosonic
string coordinate
\be \eqalign{
{X}^{\un m}(\t, \s) ~=~
\left(\matrix{
            ~ X^0 \,+\, X^3 ~~ &~~X^1 \,-\, iX^2  ~ \cr
            ~ X^1 \,+\, iX^2 ~~ &~~ ~~X^0 \,-\, X^3 ~
\cr} \right) ~~~,
}  \label{eq:52}  \ee
and fermionic string coordinates are defined by
\be \eqalign{
{{\Theta}}^{\m}{}^{i} (\t, \s)  ~~~~~
&i = 1 ,..., n_L ~~~({\rm {4D}}-{\rm {Weyl ~spinor}} )  ~~~ , \cr
{ {\Theta}}^{\m}{}^{i'} (\t, \s)  ~~~~~
&i' = 1 ,..., n_R ~~~({\rm {4D}}-{\rm {Weyl ~spinor}} ) ~~~.
}  \label{eq:53}  \ee
To complete the definition of this model, we finally define
${\hat G}_{\un A \un B \un C}$ by
  \be \eqalign{ {~~~}
{\hat G}_{\un A \un B \un C}
  =  i \frac 12 C_{\a \g} C_{\dot \b \dot \g}  \left\{ \matrix{
~
{\d}_{i} {}^{j} ~~ &~~: if~ {\un A}
= \a~ i ~,~ {\un B} = {\dot \b}~ j ~,~ {\un C} = \g \dot \g ~  \cr
        ~ $~~$ ~~ &~~ {\rm or~any~even~permutation,} ~~~~~~~~~ \cr
~ -  {\d}_{i}{}^{j}
  ~~ &~~: {\rm for~any~odd~permutation,} ~~~~~~~ \cr
~ -  {\d}_{i'} {}^{j'} ~~ &~~: if~ {\un A}
= \a~ i' ~,~ {\un B} = {\dot \b}~ j' ~,~ {\un C} = \g \dot \g ~ \cr
        ~ $~~$ ~~ &~~ {\rm or~any~even~permutation,} ~~~~~~~~~ \cr
~   {\d}_{i' }{}^{j'}
  ~~ &~~: {\rm for~any~odd~permutation,} ~~~~~~~ \cr
0 ~~ &~~:~~{\rm otherwise.}   ~~~~~~~~~~~~~~~~~~~~~~~~~
\cr } \right\}    ~~~.
}  \label{eq:54}  \ee
Remarkably, there exist ($n_L + n_R$) $\k$-supersymmetries for this action.
The elements in the first row of table IV seem to be related to the cases
of $n_L$ = 8 and $n_R$ = 0, $n_L$ = 4 and $n_R$ = 4 and $n_L$ = 4 and $n_R$
= 4, respective.  As such, these ($n_L, \, n_R$) $\k$-supersymmetric
4D GS models bare a striking similarity to the ($p,\,q$) NSR heterotic
models.  Using the ``(SUSY$)^2$--philosophy'' \cite{R31} (also known as
``superembeddings'' \cite{R32}), it is possible that there is a
correspondence between these distinct constructions of string theories.

A subsequent investigation by Siegel \cite{R33} proposed that this
construction can be interpreted fully in string theory as a part of
a new type of closed 4D GS string with ($n_L + n_R$)-extended target
space supersymmetry and further found\footnote{Berkovits and Siegel
also used this approach to analyze 4D effective actions \cite{R34}.}
that using these notions for some low values of $n_L$ and $n_R$ it
was possible to derive new supersymmetry  multiplets.  This has the
obvious additional implication, there may be even more ``animals''
in the 4D, $N$ = 8 supergravity zoo!

\section {Rejoining the Clash: de Sitter vs. SUSY}

~~~~The first challenge of describing how de Sitter background geometries
occurred in the context of extended supergravity models was overcome
by showing that such constructions exist.  However, it cannot be over
emphasized that in the de Sitter phase, all supersymmetries are
spontaneously broken.  The field is confronting a second and similar
such challenge presently in the context of superstring/M-theory.  This
time the challenge seems to be more than simply a theoretical one.

Experimental results \cite{R4,R5} point to the idea that indeed there
is a small positive cosmological constant (de Sitter geometry) in Nature.
Although there has not yet appeared creditable evidence for the
existence of supersymmetry, its presence still remains as an attractive
one to resolve the naturalness problem of the standard model.

De Sitter geometries present a considerable challenge to theoretical
physics.  Firstly, outside the context of supersymmetrical systems,
de Sitter geometries seem to lead to some problems in relativistic
quantum field theories.  As we have seen, globally supersymmetric
systems absolutely forbid de Sitter geometries.  Locally supersymmetric
systems allow them only to arise as spontaneously broken phases.
In the context of string and superstring theory, the problem of
understanding these systems in the presence of either a de Sitter
or anti-de Sitter backgrounds is largely unsolved.    Thus, superstring
theory today is in much the same situation as extended supergravity
theory in the early eighties.  The question becomes one of whether
history will repeat itself?   In this final section, we will report
on new works that suggests that the question will be answered in the
affirmative.  We will also indicate a possible avenue of approaching
this problem by use of studies of NSR non-linear $\s$-models.

The topic of string or superstring theory on a spacetime background
of constant curvature is a notoriously difficult one.  Very little
exists in the literature on this.  There has been some discussion
to the effect that on such a background, the notion of the critical
dimension of the string no longer applies.  Another work \cite{R35}
has reported on some progress in the case of the bosonic string.  For
the case of the superstring there is even less known with certainty.

To date the strongest evidence supporting the presence of de Sitter
backgrounds in superstring theory has been suggested in the work
of \cite{R36}.  Typically in theories with a spontaneously broken
symmetry, after the symmetry breaking there remains some type of
relations on some masses in the theory.  For example, in electroweak
breaking this relation is represent by the weak mixing angle.  In
the work of Herdeiro, Hiran and Kallosh \cite{R36} it has been
reported that there are such relations possible for the supertrace
of a mass matrix for the superstring level by level.

In the previous chapter, we discussed reasons for believing that
there exists at least three distinct 4D supergravity theories which
possess eight supercharges.  One of these was the 4D, $N$ = 8B
theory.  It is interesting to note that there has recently been
presented arguments by Berglund, H\" ubsch and Minic \cite{R37}
to the effect that certain warped compactification of the IIB
string theory exist wherein a spontaneously broken supersymmetry
phase with de Sitter geometry can occur.  Furthermore in their
models, the phenomenon discussed in \cite{R24} is also present
but with a naturally small cosmological constant arising.

Some years ago, it was shown \cite{R38} how to write world
sheet NSR $\s$-models which describe all the 4D, $N$ = 4 massless
modes of the 4D, $N$ = 4 SO(44) heterotic string and wherein all
bosonic condensates are explicitly represented in a (1,0) world sheet
action (see \cite{R38} for notation and conventions)
\be  { \eqalign{ {~~~}
S_{cond} ~=~ \int d^2 \s d \zeta ~ E^{-1} \Big\{ \, { 1 \over 4
\pi {\a}' }  \large[~ &  i\, (\,  g_{ \un m \un  n}(X) ~+~ B_{\un
m \un n}(X) \,) \, (\nabla_+ {X}^{\un m}) (\nabla_{\mm}
{  X}^{\un n})
{~~~~~~~~~~~~~~~~~~~~~~~} \cr
  &+\, \Phi(X) {\S}^+  ~-~ \fracm 12 \eta_-{}^{\hat I }
\nabla_+ \eta_-{}^{\hat I} \cr
~&+~ i{  { 1 \over 2 }} [~ {L_+}^{\hat \a} ( {L_{\mm}}^{\hat \a}
+ 2 {l_{\mm}}^{\hat \a} ) ~ + ~ {\L}_+ {}^{\mm}   { \tilde L
}_{\mm}{}^{\hat \a}{ \tilde L }_{\mm}{}^{\hat \a} ~ ] \cr
{~~~} &+\,{ 1 \over 2 \sqrt {2 \pi {\a}'} }~ (
\nabla_+ {  X}^{\un m}) \eta_-{}^{\hat I } A_{ \un m} {}_{ \hat I
\hat J }(X)
\eta_-{}^{\hat J }   ~\Big\} ~~ ,  }
}    \label{eq:55}  \ee
where the various quantities appearing in the action are defined
by
\be \eqalign{ {~~~~~}
&{L_+}^{\hat \a} \equiv \nabla_+ {\Phi_{L}}^{ \hat \a } ~~, ~~
{L_{\mm}}^{\hat \a} \equiv  \nabla_{\mm} {\Phi_{L}}^{\hat \a } ~~~,
  ~~ { \tilde L }_{\mm}{}^{\hat \a} \equiv {L_{\mm}}^{\hat \a} +
   l_{\mm} {}^{\hat \a }  ~~, \cr
& l_{\mm} {}^{\hat \a } \equiv { 1 \over \sqrt {2 \pi {\a}'} }~
  (\nabla_{\mm} { X}^{\un m})
A_{\un m} {}^{\hat \a}(X) + i \eta_-{}^{\hat I }\eta_-{}^{\hat J }
{\Phi}_{ \hat I \hat J } {}^{\hat \a}(X) ~~.
}  \label{eq:56}  \ee
with $\hat I$ = 1, $\dots$ , 44 and $\hat \a$ = 1, $\dots$ , 6.
The condensates are those of a 4D, $N$ = 4 supergravity multiplet
$(g_{ \un m \un  n}  ,\ A_{\un m} {}^{\hat \a}, \, B_{ \un m \un n},
\, \Phi \, ) $ and a 4D, $N$ = 4 super Yang-Mills matter multiplet
for the gauge group SO(44) $(A_{\un m} {}_{\hat I \hat  J } , \,
{\Phi}_{ \hat I \hat J } {}^{\hat \a}  \, )$.  The bosonic
supergravity condensates correspond to the theory in the work of
\cite{R11}.

 From the 4D, $N$ = 4 supergravity results, we know that the de Sitter
background cannot arise from the model above.  We first implement, on
the $\s$-model, the modification that corresponds to the
Nicolai-Townsend mapping as described in (\ref{eq:27}).  On the
$\s$-model side this is accomplished as
\be   \eqalign{ {~~~~~~}
i \int d^2 \s d \zeta ~ E^{-1} \, & \, B_{\un m \un n} \, (\nabla_+
{ X}{}^{\un m}) (\nabla_{\mm}  {  X}{}^{\un n}(X))  ~~\to~~  \cr
&i \int d^2 \s d \zeta ~ E^{-1} \, \e_{\un m \un n \un r \un s}
(\, \nabla^{\un m} B({\Hat X})  \,) \, (\nabla_+ {  {\Hat
X}}{}^{\un n}) (\nabla_{\mm}  { {\Hat X}} {}^{\un r})  (\pa_{y}  {
{\Hat X}} {}^{\un s}) ~~~, }  \label{eq:57}  \ee
via use of the standard extension of the WZNW term by the introduction
of an extra bosonic coordinate $y$ so that ${ {\Hat X}}= {{\Hat
X}}(y,\, \z,\, \t , \, \s)$ with ${ {\Hat X}}(y = 1, \,\z,\, \t ,
\, \s) =  {\Hat X}(\z,\, \t , \, \s)$ and ${ {\Hat X}}(y = 0, \,\z,\,
\t , \, \s) = 0$.

The action in (\ref{eq:55}) also contains six leftons ${\Phi_{L}}^{
\hat \a }$ and this is indicative of the [U(1)$]^6$ gauge group that
occurs for the gravi-photons whose condensates are represented by
$A_{\un m} {}^{\hat \a}$ above.  These six U(1) gauge fields must
be replaced by six SU(2) $\otimes$ SU(2) gauge fields.  However,
the way to do this is to replace the six left-moving U(1) currents
(carried on the world sheet by ${\Phi_{L}}^{\hat \a }$) by
SU(2) $\otimes$ SU(2) currents.  This calls for a Lagrangian
non-Abelian bosonization of the left-moving ${\Phi_{L}}^{\hat \a
}$ superfields.  The accomplishment of this uses the discussion given
in \cite{R39}.  There are also the SO(44) currents on the world sheet
that are carried by the forty-four spinorial superfields
$\eta_-{}^{\hat I }$.  Thus the part of the Lagrangian that
carries the currents associated with the internal symmetries
of the 4D, $N$ = 4 supergravity model is
given by
\be  { \eqalign{ {~~~}
{\cal L}_{current} ~=~ &-~ \fracm 12 \eta_-{}^{\hat I }
\nabla_+ \eta_-{}^{\hat I} ~+~ i{  { 1 \over 2 }} [~ {L_+}^{\hat \a} (
{L_{\mm}}^{\hat \a}  + 2 {l_{\mm}}^{\hat \a} ) ~ + ~ {\L}_+ {}^{\mm}
{ \tilde L }_{\mm}{}^{\hat \a}{ \tilde L }_{\mm}{}^{\hat \a} ~ ] \cr
{~~~} &+\,{ 1 \over 2 \sqrt {2 \pi {\a}'} } ~(
\nabla_+ {  X}^{\un m}) \eta_-{}^{\hat I } A_{ \un m} {}_{ \hat I
\hat J }(X)\eta_-{}^{\hat J }  ~~~ ,  }
}    \label{eq:58}  \ee
The world-sheet currents to which the gravi-photons $A_{\un m}
{}^{\hat \a}(X)$ couple are purely left-moving target space
internal currents and right-moving target space spacetime currents
while the currents to which the SO(44)-gluons $A_{ \un m} {}_{ \hat I
\hat J }(X)$ couple are purely right-moving target space internal
currents and left-moving target space spacetime currents.  Finally,
the SO(44)-scalar-gluons ${\Phi}_{ \hat I \hat J } {}^{\hat \a}(X)$
couple to the product of right-moving target space internal
currents times left-moving target space internal currents.

Following the results in \cite{R39} we are going to bosonize
the left-moving ${\Phi_{L}}^{\hat \a}$ superfields and as well as
the right-moving $\eta_-{}^{\hat I }$ superfields.  This will
amount to a replacement of these variables according to
\be  \eqalign{
{\Phi_{L}}^{\hat \a} &\to~ exp\Big[ \,i \varphi_L^{\hat \a} \,
t_{\hat\a} \, \Big] ~~~,  ~~~
\eta_-{}^{\hat I } ~\to~ exp\Big[ \, i \varphi_R^{\hat I} \,t_{\hat I}
\, \Big] ~~~.  }    \label{eq:59}  \ee
where both $\varphi_L^{\hat \a}$ and $\varphi_R^{\hat I}$ are
bosonic (1,0) superfields.  In these expressions, the quantities
$t_{\hat\a}$ and $t_{\hat I}$ respectively denote matrices representing
the groups SU(2) $\otimes$ SU(2) and SO(44).  This choice fixes the
group dimension $d_G$ and the dual co-exeter number $c_2$ associated
with Kac-Moody algebras.  This does not fix the level numbers,
however.  Insight into this can be gained by considering the anomaly
coefficients ($\n_L, \, \n_R$) \cite{R38} associated with these groups.

If $\varphi_L^{\hat \a}$ and $\varphi_R^{\hat I}$ are the coordinates
for two respective groups $G_L$ and $G_R$, then their anomaly
coefficients are given by (see also the appendix)
\begin{center}
\centerline{{\bf Anomaly Coefficients}} \vskip.1cm
\renewcommand\arraystretch{1.2}
\begin{tabular}{|c|c|c| }\hline
$~~~~$ & $~~~$ $\nu_L$ $~~~$  & $~~~$ $\nu_R$ $~~~$
   \\ \hline\hline
$ {\Phi_R}^{\hat a }  $ & $ 0 $  &  $  d_{G_R}  \Big[\,
  1 + { c_2(G_R) \over 2k_R } \, \Big]^{-1} $    \\ \hline $
{\Phi_L}^{
\hat \a }  $ & $  d_{G_L} \Big[\, 1 + { c_2 (G_R) \over 2k_L }
\, \Big]^{-1} + \fracm 12 d_{G_L} $ & $ 0 $   \\ \hline
\end{tabular}
\end{center}
\vskip.1cm
\centerline{{Table V}}
and for $G_L$ = SU${}_{k_1}$(2) $\otimes$ SU${}_{k_2}$(2) and $G_R$
= SO${}_{k_3}$(44) this leads to
\be  \eqalign{
\nu_L &=~ 3 \, \Big\{ \, 1 ~+~ \Big[\,
  1 + { 1 \over k_1 } \, \Big]^{-1} ~+~ \Big[\,
  1 + { 1 \over k_2 } \, \Big]^{-1}
  \, \Big\} ~~~, \cr ~~~ \nu_R &=~ 22 \cdot 43 \,
  \Big[\,  1 + { 42 \over k_3 } \, \Big]^{-1}~~~.
   }    \label{eq:60}  \ee
Before the ``shift'' of the gauge group from [U(1)$]^6$ to SU(2)
$\otimes$ SU(2), the fields of (\ref{eq:58}) possessed anomaly
coefficients of $\nu_L$ = 9 and $\nu_R$ = 22.  So the condition
for there to be no world sheet anomalies introduced in the
$\s$-model by the gauge group shift is
\be  \eqalign{
3 &=~ \Big\{ \, 1 ~+~ \Big[\, { k_1 \over {k_1 + 1 }} \, \Big]
~+~ \Big[\,  { k_2 \over {k_2 + 1 }} \, \Big] \, \Big\} ~~~,
  ~~~ 1 ~=~  43 \, \Big[\, { k_3 \over  {k_3 + 42 }} \, \Big] ~~~.
   }    \label{eq:61}  \ee
for some integers $k_1$, $k_2$ and $k_3$.  From the first equation,
it is seen that the condition is equivalent to $k_1$ + $k_2$ + 2 = 0.
Since level numbers are normally considered to be positive integers,
there are no solutions to this equation\footnote{There is a curiosity
about this equation.  As $k_1,\, k_2 \to \infty$, this equation is
satisfied.  One is \newline ${~~~~~~}$thus led to speculate on the
possibility of an infinite level Kac-Moody model.}.
Owing to the sake of simplicity we might as well set $k_1$ = $k_2$ =
1.  For the the second of these equations, there is a solution for
$k_3$ = 1.  This is reassuring because this means that the matter
sector of the chirally non-Abelian bosonized $\s$-model is equivalent
to the original fermionic description for this choice of the level
number.

To complete our discussion of the chiral non-Abelian bosonization
of ${\cal L}_{current}$, we will give the explicit form of its
replacement below.  However prior to doing so, there is a small
matter of some notation to clarify.  In (\ref{eq:58}) the index
${\hat I}$ takes on values 1, $\dots$ , 44.  Thus the
counting of a condensate which possesses a pair of these indices
(anti-symmetrized) yields a total of 946.  This, of course, is
the dimension of the adjoint representation of SO(44).
After the bosonization, it is more convenient to introduce
an index that takes its values in the adjoint of SO(44) and
thus the range of this index is from 1, $\dots$ , 946.  We
will denote this index also by the symbol ${\hat I}$.

Now finally for the explicit expression for the replacement
Lagrangian we have
\be  \eqalign{ {~~~}
{\cal L}_{current} ~~~\to~~~ {\cal L}_{current}^{NAB}
}    \label{eq:62}  \ee
with the latter Lagrangian given by
\be \eqalign{ {~~~~}
{\cal L}_{current}^{NAB} ~=~  &-  ~ (\, L_{\mm}^{\hat \a} \,+\,
\G_{\mm}^{\hat \a} \,) (\, L_+^{\hat\a} \,-\, \L_+ {}^{\mm}
(L_{\mm}^{\hat\a} \,+\, \G_{\mm}^{\hat\a}))  \,-\, L_+^{\hat\a}
\G_{\mm}^{\hat\a} \cr
&-~  (\, R_+^{\hat I} \,+\, 2 \G_+^{\hat I} \,) R_{\mm}^{\hat I}
\,\,+\,\, i  \L_{\mm}{}^{\pp} (\,R_+^{\hat I} \,+\, \G_+^{\hat I} )
\de_+ (R_+^{\hat I} \,+\, \G_+^{\hat I}) \cr
& +~ i {2\over3} f^{\hat I \hat J \hat K} \L_{\mm}{}^{\pp}
(R_+^{\hat I} \,+\, \G_+^{\hat I}) (R_+^{\hat J} \,+\, \G_+^{\hat J})
(R_+^{\hat K} \,-\ha \, \G_+^{\hat K}  )\cr
&-~ \int_{0}^1 d  y \, {\rm {Tr}}\{ {d \tilde L \over dy} \tilde
L^{-1} \, [\, \de_{\mm}((\de_+\tilde L) \tilde L^{-1} )
\,-\,  \de_{+}((\de_{\mm}\tilde L )\tilde L^{-1} ) ~]~ \Big\}
\cr
&+~ \int_{0}^1 d  y \,{\rm {Tr}}\{ {d \tilde R \over dy} \tilde
R^{-1}\, [\, \de_{\mm}((\de_+\tilde R \,)\, \tilde R^{-1} )
\,-\,  \de_{+}((\de_{\mm}\tilde R) \tilde R^{-1} )~] ~ \}  \cr
&-~ 4 \Phi^{\a \hat I} \Sc R_{\mm}^{\hat I}  \Sc L_+ ^{\hat \a}
~+ ~ 4 \L_+{}^{\mm} (M^{-1})^{\hat I\hat K}\Phi^{\hat \a \hat I}
\Phi^{\hat \a \hat J} \S_{\mm}^{\hat J} \S_{\mm}^{\hat K} \cr
&+~ i 4 \, \L_{\mm}{}^{\pp} \Phi^{\hat \a \hat I} \Sc L_+^{\hat
\a} \nabla_+ ( \Phi^{\hat \b \hat I} \Sc L _+^{\hat \b} ) \cr
&-~  i4 \,  \L_{\mm}{}^{\pp} f^{\hat I \hat J \hat K} ( \G_+^{\hat
I} \,+\, \frac 23  \Phi^{\hat \a \hat I } \Sc L_+^{\hat \a})
\Phi^{\hat \b \hat J} \Sc L_+^{\hat \b} \Phi^{\hat \g \hat K}
\Sc L_+^{\hat \g}
}  \label{eq:63} \ee
This Lagrangian is quite complicated and the definitions of the
various objects that appear in it can be found below.
$$\eqalign{
L_m^{\hat \a} t^{\hat\a} ~\equiv~ - i (\de_m L)L^{-1}  ~~~, ~~~
R_m^{\hat I}t^{\hat I} ~\equiv~ - i (\de_m R)R^{-1}
~~~,  }   $$
\be \eqalign{ {~~~~~~}
\Sc L_+ ^{\hat \a} ~=~ &\  L _+ ^{\hat \a} - \L _+{}^{\mm} (
L_{\mm}^{\hat \a} \,+\, \G_{\mm}^{\hat\a}) ~~~,\cr
\Sc R_{\mm} ^{\hat I} ~=~ &\ R _{\mm} ^{\hat I} - i [\L _{\mm}
{}^{\pp} \de _+ (R_+ ^{\hat I} + \G_+^{\hat I}) + \ha (\de _+
\L_{\mm}{}^{\pp}) (R_+ ^{\hat I} + \G _+ ^{\hat I}) \cr
&\,  +~ \ha \L_{\mm}{}^{\pp} f ^{\hat I \hat J \hat K}
(R_+ ^{\hat J} + \G_+ ^{\hat J}) (R_+ ^{\hat K} - \G_+^{\hat
K}) ] ~~~,\cr
\S_{\mm}^{\hat I} ~=~ &\ \Sc R_{\mm}^{\hat I} \,-\, 2 i [
\L_{\mm}{}^{\pp} \nabla_+ (\Phi^{\hat \b \hat I} \Sc L_+^{
\hat \b} ) \,+\, \ha (\nabla_+ \L_{\mm}{}^{\pp} ) \Phi^{\hat
\b \hat I} \Sc L_+^{\hat \b} \cr
&\, -~\L_{\mm}{}^{\pp} f^{\hat I \hat J \hat K} \Phi^{\hat
\b \hat J}\Sc L_+^{\hat \b} (\G_+^{\hat K}+\Phi^{\hat \g
\hat K}\Sc L_+^{\hat \g})] ~~~,\cr
( M )^{\hat I \hat J} ~=~ & \ \d^{\hat I \hat J} \,-\, 4 i \L^2
\Phi^{\hat \a \hat I} \Phi^{\hat \a \hat J} ~~~, \cr
\G_+^{\hat I} ~=~ & \,{ 1 \over 2 \sqrt {2 \pi {\a}'} }~(
\nabla_+ {  X}^{\un m}) \, A_{ \un m} {}^{ \hat I}(X) ~~~, \cr
\G_{\mm}^{\hat \a} ~=~ &  \,{ 1 \over 2 \sqrt {2 \pi {\a}'}
}~(\nabla_{\mm} {  X}^{\un m}) \, A_{ \un m} {}^{ \hat \a}(X)
~~~, \cr
\Phi^{\hat \a \hat I} ~=~ & \Phi^{\hat \a \hat I} (X)
~~~.}
\label{eq:64} \ee

The consequences of the analysis it clear.  On the 4D, $N$ = 4
supergravity side, the gauging of both SU(2) $\otimes$ SU(2) groups
(with coupling constants $g_1 = - g_2$) using gravi-photons is required
to obtain a model that possesses a de Sitter phase.  On the
$\s$-model side, this same gauging necessarily leads to an anomaly
being present.  A potential appears on the side of the supergravity
model and it is somehow related to the appearance of an anomaly
on the side of the $\s$-model.  This suggests that the supergravity
model, if it can be embedded within a $\s$-model, might correspond
to a deformation of the $\s$-model.  Once we know that the ordinary
$\s$-model which is most closely related to the Freedman-Schwarz model
is anomalous, then we can come to the end of how to proceed further
within the confines of the present state-of-the-art.  In particular,
the matter of changing the modulus choice becomes moot.

We are, however, left free to speculate.  If a quantum theory is
anomalous, past experience has taught us that this is a signal
that there may be extra terms, extra degrees of freedom, etc. that
must be introduced to augment the original theory.  Perhaps this
is what is required in the present circumstance also.   Looking back
at the failure of the left-anomaly cancellation condition, we note
that if the lefton current group is changed from SU${}_1$(2)
$\times$ SU${}_1$(2) $\times$ U(1) or even SU${}_1$(2) $\times$
SU${}_1$(2) $\times$ SU${}_1$(1), then the condition for left
anomaly-freedom is satisfied.  From the 4D, $N$ = 4 supergravity
side, we know that there are no fundamental massless excitations
associated with the introduction of the final group.  Still its
presences signals the possibility to introduce even more $\b$-functions
that might correspond to composite functions of the original ones.
So there seems to be at least a hope that some augmentation might work.
Of course, there is still the more stringent test of whether a
lefton group of the form of SU${}_1$(2) $\times$ SU${}_1$(2) $\times$ U(1)
or SU${}_1$(2) $\times$  SU${}_1$(2) $\times$ SU${}_1$(1) is acceptable.

The strongest evidence to support the possibility that an augmented
action can be found is the suggestion of the ``stringy Zeeman effect''
\cite{R27,R36}.  A completely successful realization of the tantalizing
hint coming from the stringy Zeeman effect will demand this.  Thus, a
model \cite{R18} found nineteen years ago may have yet another role
as a laboratory in which to unravel the mystery of treating superstrings
in a de Sitter space background.

\section {Conclusion}

~~~~A universe in which there exists supersymmetry simultaneously
with a background geometry of the de Sitter variety presents
real challenges to superstring theory.  A challenge like this
is extremely healthy in this author's opinion.  It will clearly
demand an advance of the state-of-the-art in superstring
theory and perhaps more generally in field theory.  It is very
likely to force a better understanding of the meaning of the
word ``geometry'' in the superstring arena.

With this presentation we hope to have achieved a few simple
goals.  These were;

  ${~~~~~}$ (a.) to have provided as complete as possible a review
       of the \newline ${~~~~~~~~~~~~~~~\,}$ early literature on the
       problem of the admissibility of de \newline ${~~~~~~~~~~~~~~~\,}$
       Sitter space background in the confines of theories with
       \newline ${~~~~~~~~~~~~~~~\,}$ local supersymmetry,

${~~~~~}$ (b.) to have illustrated, in some detail, how a 1984 work
        pro- \newline ${~~~~~~~~~~~~~~~\,}$ vided a paradigm to by-pass
        a general no-go theorem rul- \newline ${~~~~~~~~~~~~~~~\,}$ ing
        out de Sitter space background in confines of super- \newline
        ${~~~~~~~~~~~~~~~\,}$ gravity theories,

${~~~~~}$ (c.) to make the community aware of some long standing
         open \newline ${~~~~~~~~~~~~~~~\,}$ issues related to the
         cosmological constant in supergravity \newline
         ${~~~~~~~~~~~~~~~\,}$ theories,

${~~~~~}$ (d.) to note some encouraging points of recent literature
          that \newline ${~~~~~~~~~~~~~~~\,}$ suggest the problem of
          the admissibility of de Sitter space \newline
         ${~~~~~~~~~~~~~~~\,}$  background in the confines of
         superstring theories will re- \newline ${~~~~~~~~~~~~~~~\,}$
         peat the past pattern of this problem within the confines
         \newline ${~~~~~~~~~~~~~~~\,}$ of supergravity theory and

${~~~~~}$ (e.) to provide a detailed analysis from the point of the
          view \newline ${~~~~~~~~~~~~~~~\,}$ of the combined world
          sheet $\s$-model/4D, $N$ = 4 supergra- \newline
          ${~~~~~~~~~~~~~~~\,}$ vity system as what are
          the hurdles that must be surmount- \newline
          ${~~~~~~~~~~~~~~~\,}$ ed if the de Sitter phase, known
          to exist in 4D, $N$-extended  \newline ${~~~~~~~~~~~~~~~\,}$
          supergravity (equivalent to higher D SG theories)
          models,  \newline ${~~~~~~~~~~~~~~~\,}$ is to be extended
          for superstrings.

Finally and of course, we wish to celebrate the initial presentation
of the theory of supergravity and hope that Nature is aware of our efforts
(as one colleague in the field has been heard to say).

${~~~}$ \newline
${~~~~~~~~~}$``{\it {Let only geometers
enter here.}}'' \newline
${~~~~~~~\,~~}$ \newline
${~~~~~~~~~~~}$ -- Copernicus

\newpage
\noindent
{\bf {Acknowledgment}} \newline \noindent
${~~~~}$We wish to acknowledge the organizers of the
``Supergravity\@25'' conference for their invitation to make
a presentation.  We also wish to acknowledge our hosts (Hendrik
Geyer and Joao Rodrigues) at the Fourteenth Chris Engelbrecht
Summer School in Theoretical Physics and as well the Stellenbosch
Institute for Advanced Study (STIAS, Dir. Bernard Lategan) workshop
on String Theory and Quantum Gravity, held in Stellenbosch, South 
Africa during Jan. 23 to  Feb. 22, 2002.  The lively atmosphere of 
the school and workshop provided a stimulating environment in
which to complete the writing this extended version of my Stony
Brook presentation.  Finally, it is an honor to contribute this
paper as the first STIAS publication.

\noindent{{\bf {Appendix: Groups, Dimensions and Dual Co-exeter
Numbers}}}

~~~~In obtaining the results in (\ref{eq:60}), we have made use
of some results that certainly exist in the prior physics literature.
We present them here for the convenience of the reader. For a group
with matrices $t_{\hat a }$ that faithfully represent its
generators , we have used the following definitions.
$\hat a = 1 , \dots ,d_G~$, $ [ t_{\hat a },~t_{\hat b } ] = i
f_{\hat a \hat b}{}^{\hat c} t_{\hat c }~$, $f_{\hat a \hat b \hat c}
f^{\hat a \hat b}{}_{\hat d} = c_2 {\d}_{\hat c \hat d }~$ and $ Tr (
t_{\hat a }  t_{\hat b } ) = 2k \d_{\hat a\hat b}$.

\begin{center}
\centerline{{\bf {Values for}} ${d_G}$ {\bf {and}} $ {c_2}$ }
\vskip.1in
\renewcommand\arraystretch{1.2}
\begin{tabular}{|c|c|c| }\hline
$~~~~~$ Group $~~~~~$ & $~~~~~~~d_G~~~~~~~$  & $~~~~~~c_2~~~~~~$
   \\ \hline\hline
SU(n) & $ n^2 \,-\, 1 $  &  $  n $    \\ \hline
SO(2n \,+\, 1) & $ n \, ( 2n \,+\, 1) $  &  $  4 n \,-\, 2$  \\ \hline
Sp(n) & $ n \, ( 2n \,+\, 1) $  &  $  2 n \,+\, 2$    \\ \hline
SO(2n) & $ n \, ( 2n \,-\, 1) $  &  $  4 n \,-\, 4$  \\ \hline
G${}_2$ & $ 10 $  &  $  8 $  \\ \hline
F${}_2$ & $ 40 $  &  $  18 $  \\ \hline
E${}_6$ & $ 78 $  &  $  24 $  \\ \hline
E${}_7$ & $ 133 $  &  $  36 $  \\ \hline
E${}_8$ & $ 248 $  &  $  60 $  \\ \hline
\end{tabular}
\end{center}
\vskip.1in
\centerline{{Table VI}}

\end{document}